\documentclass[twocolumn,twocolappendix]{aastex631}

\usepackage{physics}
\usepackage{amsmath}
\usepackage{tabu}
\usepackage{booktabs}
\usepackage{subfigure}
\usepackage{hyperref}
\usepackage{ulem}
\usepackage{enumerate}
\usepackage{graphicx}
\usepackage{CJK}

\newcommand{\sect}{Section~}

\newcommand{\figu}{Figure~}
\newcommand{\figus}{Figures~}
\newcommand{\tab}{Table~}

\newcommand{\eq}{Equation~}
\newcommand{\eqs}{Equations~}
\newcommand{\athenapp}{\texttt{Athena++}}
\newcommand{\athenak}{\texttt{AthenaK}}

\shorttitle{Supernova Remnants}
\shortauthors{Guo et al.}
\graphicspath{{./}{figures/}}

\begin{document}
\begin{CJK*}{UTF8}{gbsn}

\title{Evolution of Supernova Remnants in a Cloudy Multiphase Interstellar Medium}

\correspondingauthor{Minghao Guo}
\email{mhguo@princeton.edu}

\author[0000-0002-3680-5420]{Minghao Guo (郭明浩)}
\affiliation{Department of Astrophysical Sciences, Princeton University, Princeton, NJ 08544, USA}

\author[0000-0003-2896-3725]{Chang-Goo Kim}
\affiliation{Department of Astrophysical Sciences, Princeton University, Princeton, NJ 08544, USA}

\author[0000-0001-5603-1832]{James M. Stone}
\affiliation{School of Natural Sciences, Institute for Advanced Study, 1 Einstein Drive, Princeton, NJ 08540, USA}
\affiliation{Department of Astrophysical Sciences, Princeton University, Princeton, NJ 08544, USA}

\begin{abstract}
We investigate the evolution of supernova remnants (SNRs) in a two-phase cloudy medium by performing a series of high-resolution (up to $\Delta x\approx0.01\,\mathrm{pc}$), 3D hydrodynamical simulations including radiative cooling and thermal conduction. We aim to reach a resolution that directly captures the shock-cloud interactions for the majority of the clouds initialized by the saturation of thermal instability. In comparison to the SNR in a uniform medium with the volume filling warm medium, the SNR expands similarly (following $\propto t^{2/5}$) but sweeps up more mass as the cold clouds contribute before shocks in the warm medium become radiative. However, the SNR in a cloudy medium continuously loses energy after shocks toward the cold clouds cool, resulting in less hot gas mass, thermal energy, and terminal momentum. Thermal conduction has little effect on the dynamics of the SNR but smooths the morphology and modifies the internal structure by increasing the density of hot gas by a factor of $\sim 3-5$. The simulation results are not fully consistent with many previous 1D models describing the SNR in a cloudy medium including a mass loading term. By direct measurement in the simulations, we find that, apart from the mass source, the energy sink is also important with a spatially flat cooling rate $\dot{e}\propto t^{-11/5}$. As an illustration, we show an example 1D model including both mass source and energy sink terms (in addition to the radiative cooling in the volume filling component) that better describes the structure of the simulated SNR.
\end{abstract}


\section{Introduction} \label{sec:intro}

Supernova (SN) explosions are believed to be crucial in shaping interstellar medium (ISM), producing the hot phase, driving turbulence, regulating the star formation rate, and launching galactic outflows (see \citealt{Cox2005ARA&A..43..337C} for a review to classical views and \citealt{Kim2017ApJ...846..133K, Kim2020ApJ...900...61K, Kim2023ApJ...946....3K} for a modern view). The explosion of SN ejects fast material with a typical mass of $M_\mathrm{ej}\sim 1$-$10\,\mathrm{M_\odot}$ and kinetic energy of $E_\mathrm{SN}\sim10^{51}\mathrm{erg}$ \citep{Draine2011piim.book.....D}. The localized injection of such high energy SN ejecta in turn launches blastwaves expanding into the surrounding medium. Earlier works have described the evolution of SNRs using spherically symmetric expansion in a uniform medium, which is often characterized into four distinct stages marked by the power law index of the spherical expansion \citep{Cox1972ApJ...178..159C, Woltjer1972ARA&A..10..129W, Cioffi1988ApJ...334..252C, Ostriker&McKee1988RvMP...60....1O, Draine2011piim.book.....D}: free expansion stage ($r\propto t$), energy-conserving stage (so-called Sedov~\citep{Sedov1959sdmm.book.....S} and Taylor~\citep{Taylor1950RSPSA.201..159T} (ST) stage; $r\propto t^{2/5}$), pressure-driven snowplow (PDS) stage ($r\propto t^{2/7}$), and momentum-conserving snowplow stage ($r\propto t^{1/4}$). 

However, the ISM is far from being uniform, consisting of multiple distinct thermal phases over a wider range of temperature and density. SNe themselves are responsible for making the ISM multiphase, by shock heating to create the hot phase with $T\sim 10^{6-7}\,\mathrm{K}$ \citep{Cox&Smith1974ApJ...189L.105C, McKee&Ostriker1977ApJ...218..148M}. In addition, the ISM is also heated by UV radiation and cosmic ray ionization, providing the major heating in the neutral medium and maintaining warm neutral medium (WNM) at $T\sim 10^4\,\mathrm{K}$ and cold neutral medium (CNM) at $T\sim 10^2\,\mathrm{K}$ \citep{Field1965ApJ...142..531F, Field1969ApJ...155L.149F, Wolfire1995ApJ...443..152W}. The molecular gas and the photoionized gas are significant constituents of the ISM, while they share similar thermal states with the CNM and WNM, respectively \citep{Kim2023ApJ...946....3K, Rathjen2021MNRAS.504.1039R}. The complexity is added by turbulence (also mainly driven by SNe), which broadens the density and temperature distributions from a single characteristic state \citep[e.g.,][]{Ostriker2001ApJ...546..980O} and continuously populates thermally unstable (thought to be transient) states \citep{Gazol2013ApJ...765...49G, Ho2024arXiv240714199H}. 
 
With the pervasive diffuse hot and warm gas filling most of the volume, the cold gas is confined in dense clouds occupying only $\sim 1\%$ of the volume of the local ISM\citep[e.g.,][]{Draine2011piim.book.....D, Kim2023ApJ...946....3K}. It is thus well motivated to consider the effect of clouds in the SNR evolution as ``impurities'' of the spherical blastwave expansions by incorporating the consequence of the shock-cloud interactions \citep[e.g.,][for a review]{Ostriker&McKee1988RvMP...60....1O}. A plethora of dedicated studies on the shock-cloud interaction (there is a separate collection of the wind-cloud interaction literature) has been conducted starting from pure hydrodynamical simulations \citep[e.g.,][]{Bedogni1990A&A...231..481B, Klein1994ApJ...420..213K, Pittard&Parkin2016MNRAS.457.4470P} to more complex problems including magnetic fields, thermal conduction (isotropic and anisotropic), turbulence, and radiative cooling \citep[e.g.,][to name a few]{MacLow1994ApJ...433..757M, Gregori1999ApJ...527L.113G, Shin2008ApJ...680..336S, Orlando2005A&A...444..505O, Goldsmith2017MNRAS.470.2427G, 2020MNRAS.499.2173B}.

The major effect often considered in such 1D impurity models is the mass loading due to cloud crushing and conductive evaporation. \citet{McKee&Ostriker1977ApJ...218..148M} argued that the cloud evaporation mass can exceed swept-up mass and even modify the expansion law during the energy-conserving stage from the usual $r\propto t^{2/5}$ to $r\propto t^{3/5}$. \citet{Cowie1981ApJ...247..908C} presented a time-dependent solution with mass, momentum, and energy source terms under a variety of assumptions about the efficiency of thermal evaporation depending on temperature, cloud size, saturation, etc. \citet{White&Long1991ApJ...373..543W} proposed a modified self-similar solution with enhanced interior density and pressure considering conductive evaporation of cold clouds. Different forms of mass loading prescriptions were considered in 1D time-dependent spherical simulations by assuming constant, Mach number dependent, and temperature-dependent mass loading rates \citep[e.g.,][]{Dyson2002A&A...390.1063D, Pittard2003A&A...401.1027P}. Recent work by \citet{Pittard2019MNRAS.488.3376P} investigated the effect of constant mass loading rate and concluded that mass-loaded SNRs attain higher radial momentum and more hot gas at shell formation but lower final momentum due to more rapid cooling and less PdV work during the PDS stage.

While well motivated, the validation of the 1D impurity models requires direct global SNR simulations in a cloudy medium with high resolution to capture the major consequences of the shock-cloud interaction. Especially, the role of radiative cooling in the turbulent mixing layers \citep{Fielding2020ApJ...894L..24F, Tan2021MNRAS.502.3179T} has recently been emphasized in the wind-cloud interaction literature (mostly in the galactic wind context), leading to qualitative changes in the results \citep{Armillotta2016MNRAS.462.4157A, Gronke2018MNRAS.480L.111G, Sparre2020MNRAS.499.4261S, Li2020MNRAS.492.1841L, Abruzzo2022ApJ...925..199A}. Namely, the interaction between the fast-moving hot gas with large clouds creates a turbulent mixing layer within which the radiative cooling is so strong that the surrounding hot gas condenses onto clouds. The net effect is then the growth of clouds with significant energy loss from the hot gas rather than simple shredding and evaporation of clouds. In a similar vein, stellar wind blown bubbles experience an efficient cooling when expanding into a turbulent medium, leading to the ``momentum-conserving''-like solution with a minimal enhancement of momentum compared to the injected wind momentum \citep{Lancaster2021ApJ...914...89L, Lancaster2021ApJ...914...90L}. It is thus important to access previous approaches using direct 3D simulations that are capable of capturing radiative cooling at the shock-cloud interaction layers.

Recent advances in hardware and software in computational astrophysics make direct 3D numerical simulations of radiative SNRs in a cloudy medium possible. About a decade ago, a number of such simulations are performed with different realizations of cloudy medium: placing cold clouds manually ~\citep{Li2015ApJ...814....4L, Slavin2017ApJ...846...77S}, turbulent molecular clouds~\citep{Walch2015MNRAS.451.2757W, Iffrig2015A&A...576A..95I}, Gaussian random field realization using log-normal density distribution \citep{Martizzi2015MNRAS.450..504M, Zhang2019MNRAS.482.1602Z}, and a saturated state of thermal instability \citep{Kim2015ApJ...802...99K, Kim2017ApJ...834...25K}. One of the main results in these simulations is that the characteristic properties of SNRs including the mass and momentum at the shell formation are still well approximated by the classical 1D results when using the mean density of the surrounding medium. These simulation results neither follow the SNR expansion solution modified by cloud evaporation \citep{McKee&Ostriker1977ApJ...218..148M} nor show significant enhancements in momentum injection before and at shell formation when a certain mass loading model is assumed \citep{Pittard2019MNRAS.488.3376P}. However, it is still unclear whether the discrepancies between 3D simulations and 1D impurity models are due to insufficient numerical resolution in the simulations to fully capture the shock-cloud interaction. On the one hand, one needs a minimal resolution of 32-64 cells per radius for a single cloud to resolve the hydrodynamical interaction and cloud ablation \citep{Pittard&Parkin2016MNRAS.457.4470P}. On the other hand, although the mass and energy exchanges in the turbulent mixing layer formally require much higher resolution (e.g., resolving the turbulent cascade down to cooling and conductive length scales), apparent convergence can achieved at lower resolutions than the formal requirements if the largest scale for turbulence is resolved \citep[e.g.,][]{Fielding2020ApJ...894L..24F, Tan2021MNRAS.502.3179T, Lancaster2024ApJ...970...18L}.

Another technological advance in the last decade is represented by the deployment of GPU-accelerated high-performance computers. When utilizing such capabilities, direct 3D simulations with about an order of magnitude higher spatial resolution are easily accessible. At the same time, adopting higher-order methods allows even higher-fidelity simulations. We use a newly developed performance portable magnetohydrodynamic code \athenak{} \citep{Stone2024arXiv240916053S} and run a series of hydrodynamical simulations of radiative SNRs in a two-phase cloudy medium (a setup similar to \citealt{Kim2015ApJ...802...99K}) at varying resolution with and without thermal conduction. The highest resolution we achieve is $1/64$ pc while we follow the global SNR evolution until the forward shock toward the WNM begins to cool at the age of $\sim 30$ kyr and radius of $\sim 20$ pc. Although our resolution is not sufficient to reach a formal resolution requirement for cooling and conductive length scales, it is high enough to resolve hydrodynamic instabilities for the majority of clouds (down to cloud size of $0.5$ pc).

In this work, we consider a condition with the ambient mean density of $n=10\,{\rm cm^{-3}}$ while we explore the effect of different realizations of a cloudy medium. We focus on the effect of cold clouds in the evolution and internal structure of the volume-filling component (i.e., the hot gas) before the shell formation in the WNM (i.e., cooling of outer shocks propagating into the WNM). This is a nominal energy-conserving stage during which the hot gas mass and radial momentum increase significantly such that the dynamical impact of SNRs is determined. We delve into the roles of the shock-cloud interaction in terms of mass loading into the hot interior by cloud ablation (including conductive evaporation) and energy loss at cloud interfaces by the development of turbulent mixing layers. Finally, we present a minimal 1D model that reproduces the 3D simulation results. We emphasize that the set of simulations presented in this work only utilizes about $1.2\times10^4$ GPU hours ($\sim 2.4\times10^3$ GPU hours for each of the highest resolution runs), implying more systematic studies are possible to motivate the development of sophisticated models.

The rest of this article is organized as follows. \sect\ref{sec:theory} summarizes the analytical theories of SNR. \sect\ref{sec:method} describes the physical model and numerical methods we adopt. In \sect\ref{sec:results} we present the results of SN in a cloudy two-phase medium and its dependence on various parameters. In \sect\ref{sec:discussion}, we discuss the implications for 1D semi-analytical models. We conclude in \sect\ref{sec:summary}.

\section{Summary of 1D, spherical SNR expansion theory} \label{sec:theory}
As a reference throughout the paper, it is useful to summarize a standard theory of 1D, spherically-symmetric SNR evolution in a uniform medium. The early theories identified four stages of evolution in which SNRs follow different expansion laws: free expansion stage ($r\propto t$), adiabatic expansion stage ($r\propto t^{2/5}$), pressure-driven snowplow (PDS) stage ($r\propto t^{2/7}$), and momentum-conserving snowplow stage ($r\propto t^{1/4}$).

During the free expansion stage when the SN ejecta expands ballistically, the SNR sweeps up more and more material. In doing so, reverse shocks develop and thermalize the interior, entering the so-called Sedov~\citep{Sedov1959sdmm.book.....S} and Taylor~\citep{Taylor1950RSPSA.201..159T} (ST) stage, a main focus of this work. In the ST stage, the explosion is well described by an adiabatic point source explosion, which is solely determined by the total energy $E_\mathrm{SN}$ and the ambient density $\rho_0$. A self-similar solution can be obtained \citep[e.g.,][]{Sedov1959sdmm.book.....S, Draine2011piim.book.....D} to give (for an adiabatic index $\gamma=5/3$)
\begin{align}
    r_\mathrm{S}&=5.88\,\mathrm{pc}\,E_{51}^{1/5}n_0^{-1/5}t_3^{2/5},\label{eq:r_s}\\
    v_\mathrm{S}&=\frac{2}{5}\frac{r_\mathrm{S}}{t}=2.35\times10^3\,\mathrm{km\,s^{-1}}\,E_{51}^{1/5}n_0^{-1/5}t_3^{-3/5},\label{eq:v_s}\\
    T_\mathrm{S}&=\frac{3}{16}\frac{\mu v_\mathrm{S}^2}{k_\mathrm{B}}=7.37\times10^7\,\mathrm{K}\,E_{51}^{2/5}n_0^{-2/5}t_3^{-6/5},\label{eq:T_s}
\end{align}
where $E_{51}=E_\mathrm{SN}/10^{51}\,\mathrm{erg}$ is the supernova expansion energy, $n_0=n/1\,\mathrm{cm^{-3}}$ is the total number density, and the mean molecular weight $\mu=0.618$ is adopted for fully ionized gas.

The ST stage ends when the radiative energy losses are no longer negligible. The postshock gas experiences rapid cooling and forms a dense shell. Using an approximation of cooling function for the postshock gas temperature range, $\Lambda(T) = 3.0\times10^{-22}{\rm erg\,s^{-1}\,cm^{3}} (T/10^6\,\mathrm{K})^{-0.7}$ \citep[slight higher than][but appropriate for our adopted cooling function]{Draine2011piim.book.....D}, one can obtain the shell formation time\footnote{We adopt the definition presented in \citet{Kim2015ApJ...802...99K}. $t_{\rm sf}$ is defined by the minimum of $t_s+t_\mathrm{cool}$, where $t_s$ is the time when the gas is shocked and $t_\mathrm{cool}$ is the cooling time of the postshock.},
\begin{align}
    t_\mathrm{sf}&=3.0\times 10^{4}\,\mathrm{yr}\,E_\mathrm{51}^{0.22}n_0^{-0.55}.
\end{align}
Plugging this into \eqs\ref{eq:r_s} and \ref{eq:v_s}, we have the size and expansion velocity at the shell formation, respectively,
\begin{align}
    r_\mathrm{sf}&=22.9\,\mathrm{pc}\,E_\mathrm{51}^{0.29}n_0^{-0.42},\\
    v_\mathrm{sf}&=299\,\mathrm{km\,s^{-1}}\,E_\mathrm{51}^{0.07}n_0^{0.13}.
\end{align}

It is worth emphasizing that the ST stage is the most dynamically important stage of SNR evolution during which SNRs build up both hot gas mass and radial momentum significantly. At the end of the ST stage, the hot gas mass (or swept up mass) and total radial momentum reach
\begin{align}
    M_{\rm sf}&= 765\,M_\odot\,E_\mathrm{51}^{0.87}n_0^{-0.26},\\
    p_{\rm sf}&= 1.47\times10^5, \,M_\odot\mathrm{km\,s^{-1}}\,E_\mathrm{51}^{0.94}n_0^{-0.13},
\end{align}
which are more than 100 and 10 times larger than the ejecta mass and momentum, respectively.
Our work focuses mainly on the detailed evolution of this stage.

As references for the analysis presented in upcoming sections, we define a reference rate of change for mass, energy, density, and energy density,
\begin{align}
    \dot{M}_\mathrm{S}\equiv{4\pi}\rho_0 v_\mathrm{S} r_\mathrm{S}^2, \\
    \dot{E}_\mathrm{S}\equiv{4\pi}\rho_0 v_\mathrm{S}^3 r_\mathrm{S}^2, \\
    \dot{\rho}_\mathrm{S}\equiv3\rho_0 v_\mathrm{S} /r_\mathrm{S}, \\
    \dot{e}_\mathrm{S}\equiv3\rho_0 v_\mathrm{S}^3 /r_\mathrm{S},
\end{align}
as well as post shock density and pressure $\rho_\mathrm{S}=4\rho_0$, $P_\mathrm{S}=3/4\rho_0v_\mathrm{S}^2$ assuming strong shock.

\begin{deluxetable*}{lcccc}
    \tablenum{1}
    \tablecaption{Definitions of phases of gas. \label{tab:gas}}
    \tablewidth{0pt}
    \tablehead{
    \colhead{Phase Name} & \colhead{Criterion} & \colhead{Label} & \colhead{Passive Scalar} 
    }
    \startdata
    Cold Neutral Medium & $T<184\,\mathrm{K}$ & CNM & $C_0$
    \\
    Unstable Neutral Medium & $184\,\mathrm{K}\leq T<5050\,\mathrm{K}$ & UNM & $C_1$ \\
    Warm Neutral Medium & $5050\,\mathrm{K}\leq T<2\times10^4\,\mathrm{K}$ & WNM & $C_2$ \\
    Initial Ejecta & \multicolumn{1}{c}{--} & EJ & $C_3$ \\[0.1cm]
    \hline\\[-0.4cm]
    Hot Gas & $T\geq2\times10^4\,\mathrm{K}$ & hot & \multicolumn{1}{c}{--} \\
    Shell Gas & $T<2\times10^4\,\mathrm{K}$ and $|v|>5\,\mathrm{km\,s^{-1}}$ & shell & \multicolumn{1}{c}{--} \\
    Supernova Remnant Gas & $T\geq2\times10^4\,\mathrm{K}$ or $|v|>5\,\mathrm{km\,s^{-1}}$ & SNR & \multicolumn{1}{c}{--} \\
    \enddata
    \tablecomments{Names assigned to the gas phases based on temperature and velocity criteria. The first four phases already exist in the initial conditions and are assigned with a passive scalar separately. The last four phases are defined during the evolution.}
\end{deluxetable*}

\section{Numerical Methods} \label{sec:method}
For direct 3D simulations of radiative SNR evolution in an inhomogeneous medium, we solve the hydrodynamic equations with radiative cooling and thermal conduction using \athenak{}~\citep{Stone2024arXiv240916053S}, a GPU-accelerated version of the \athenapp~\citep{Stone2020ApJS..249....4S} code. \athenak{} employs an unsplit Godunov algorithm to solve the equations of conservative hydrodynamics. Our simulations adopt the piecewise parabolic (PPM) reconstruction method, the Harten-Lax-van Leer-Contact (HLLC) Riemann solver, and the RK3 time integrator to achieve a high-order convergence in space and time. The source terms (radiative cooling and heating) are included using the operator splitting method.

The equations we solve are
\begin{eqnarray}
    \frac{\partial \rho}{\partial t}+\nabla \cdot(\rho \boldsymbol{v}) &=&0, \\
    \frac{\partial \rho \boldsymbol{v}}{\partial t}+\nabla \cdot\left(P\mathbf{I}+\rho \boldsymbol{v} \boldsymbol{v}\right) &=&0, \\
    \frac{\partial E}{\partial t}+\nabla \cdot\left[\left(E+P\right) \boldsymbol{v}+\boldsymbol{Q}\right] &=&-\rho\mathcal{L}, \\
    \frac{\partial (\rho C_i)}{\partial t}+\nabla \cdot(\rho \boldsymbol{v}C_i) &=&0, \label{eq:scalar}
\end{eqnarray}
where $\rho$ is the mass density, $\boldsymbol{v}$ is the velocity, $P=\rho k_\mathrm{B} T/(\mu m_\mathrm{H})$ is the gas pressure with mean molecular weight $\mu=0.618$. The mean molecular weight should vary according to the ionization state, but since we do not trace the ionization and recombination in detail, we assume that the gas is fully ionized. The total energy density is $E=e+\rho v^2/2$ where $e=P/(\gamma-1)$ the is internal energy density with the adiabatic index $\gamma=5/3$, $\boldsymbol{Q}$ is the heat flux due to thermal conduction, and $\rho\mathcal{L}$ is the net cooling rate.

In addition to the Euler equations to describe the hydrodynamic flows, we solve additional conservation equations for ``passive scalars'' (\eq\ref{eq:scalar}) to follow the different phases in the background medium separately as well as the supernova ejecta. $C_i$ is the specific density of each scalar species with $\rho C_i$ being the mass density to trace different phases of gas. We add 4 passive scalars in the simulations: $C_0$ as the CNM, $C_1$ as the UNM, $C_2$ as the WNM, and $C_3$ as the hot ejecta of the supernova. Apart from the phases that exist in the initial conditions (tagged by passive scalars), we further define the hot gas as all the gas with $T>2\times10^4\,\mathrm{K}$ and the shell gas\footnote{In the SNRs interacting with a cloudy medium, the ``shell'' gas is not only corresponding to the outer shell (or cooled forward shock) but also representing the accelerated clouds.} with $|v|>5\,\mathrm{km\,s^{-1}}$ and $T<2\times10^4\,\mathrm{K}$. The SNR gas is defined with $|v|>5\,\mathrm{km\,s^{-1}}$ or $T\geq2\times10^4\,\mathrm{K}$ to enclose all the gas that is significantly perturbed or heated. The full definition of the different phases of gas we used in our simulations is summarized in \tab\ref{tab:gas}.

The implementations of initial and boundary conditions, thermal conduction, cooling, heating, floors, and corrections are presented below. We tested our implementation of cooling and thermal conduction. As a reference run, the evolution of SNR in a uniform background is presented in Appendix \ref{app:snr_uni}.

\begin{figure}[htb]
    \centering
    \includegraphics[width=\linewidth]{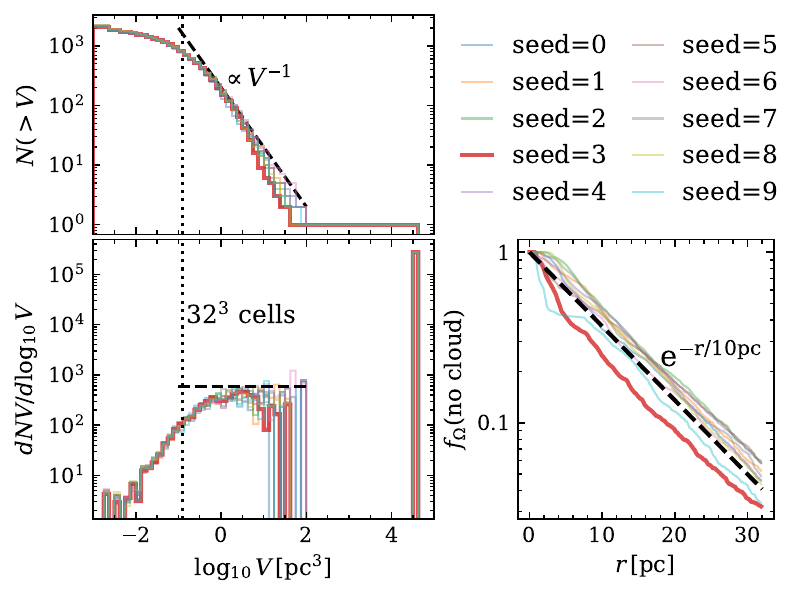}
    \caption{Left: Cumulative distribution function (top) and density distribution function of the cold cloud volume in the 10 pre-simulations of thermal instability. The vertical dotted lines mark the size of clouds resolved by $32^2$ cells. A cloud is defined to be a contiguous region of material on the cold stable thermal equilibrium. The majority of the clouds follow Zipf's Law-like distribution. A significant portion of the CNM is interconnected, forming one cohesive region through small filaments. Right: fraction of solid angle within which there is no cold neutral medium along the line of sight as a function of radius. It follows an exponential decay $f_\Omega\text{(no cloud)}\sim \exp{-r/10\mathrm{pc}}$. The dark red one (seed=3) is used as the fiducial initial condition.}
    \label{fig:cloud_dist}
\end{figure}

\subsection{Initial and Boundary Conditions}
The simulations adopt a cubic box of side $64\,\mathrm{pc}$ ($[-32,32]^3\,\mathrm{pc}$) with the center of the supernova remnant placed at the center of the simulation box. We place a uniform spherical hot ejecta with radius $r=1\,\mathrm{pc}$, mass $M_\mathrm{SN}=3\,M_\odot$ and total SN energy $E_\mathrm{SN}=10^{51}\,\mathrm{erg}$ in the form of thermal energy for simplicity. Though supernovae inject their energy mainly in the form of kinetic energy, the setup here is appropriate since the initial ejecta is small in size.

We set up a cloudy background state through pre-simulations of thermally unstable medium \citep[e.g.,][]{Kim2015ApJ...802...99K}. For the fiducial suite of pre-simulations, we begin with a uniform $n_0=10\,\mathrm{cm^{-3}}$ medium with a divergence-free Kolmogorov-like velocity perturbation and let the thermal instability develop naturally to form the two-phase medium. We set the velocity perturbation with wavelength $3.2\,\mathrm{pc}<\lambda<16\,\mathrm{pc}$ and amplitude $\sigma_v\sim 1\,\mathrm{km\,s^{-1}}$. We adopt a periodic boundary condition for all the simulations. 

In the pre-simulations of thermal instability, we observed the natural development of a two-phase medium, consisting of WNM ($T_\mathrm{WNM}\approx 6600\,\mathrm{K}$, $n_\mathrm{WNM}\approx0.86\,\mathrm{cm^{-3}}$, volume filling fraction $\sim 0.89$ and mass fraction of $\sim 0.08$) and CNM ($T_\mathrm{CNM}\approx 68\,\mathrm{K}$, $n_\mathrm{CNM}\approx84\,\mathrm{cm^{-3}}$). Additionally, a significant fraction of the medium remained in a UNM state. The distribution of the cold clouds for 10 pre-simulations is shown in \figu\ref{fig:cloud_dist}. The distribution follows a Zipf's Law-like distribution with an index of $-1$ over 3 orders of magnitude in cloud volume, yielding nearly equal mass per logarithmic bin in clump mass. This has commonly been observed in other turbulent multiphase systems \citep{Gronke2022MNRAS.511..859G, Tan2024MNRAS.527.9683T}. However, we note that despite the volume filling factor of the CNM being relatively small, a significant portion (more than $90\%$) of the CNM is interconnected, forming one cohesive region through small filaments. Our fiducial model (seed=3) has a giant cloud near the domain center with a relatively large covering fraction with $f_\Omega\text{(no cloud)}$ dropping to 50\% at 5 pc. 

\subsection{Thermal Conduction}

We consider an isotropic thermal conduction $\boldsymbol{Q}_0=-\kappa\nabla T$. We adopt the conductivity given by  ~\citet{Spitzer1962pfig.book.....S} and \citet{Parker1953ApJ...117..431P} as
\begin{equation}
\kappa=\left.
  \begin{cases}
    6\times10^{9} T_6^{5/2}\mathrm{\, erg\,cm^{-1}\,K^{-1}\,s^{-1}}, & T>6.5\times10^4 \\
    2.5\times10^5 T_6^{1/2}\mathrm{\, erg\,cm^{-1}\,K^{-1}\,s^{-1}},
    & \text{otherwise}
  \end{cases}
  \right.
\end{equation}
where $T_6\equiv T/(10^6\,\mathrm{K})$.
We adopt the saturation of heat flux following a form given in \citet{Cowie&McKee1977ApJ...211..135C},
\begin{equation}
    \boldsymbol{Q}_\mathrm{sat}=-5\phi_sPc_{s,\,\mathrm{iso}}\frac{\nabla T}{|\nabla T|}=-\frac{3}{2}Pc_{s,\,\mathrm{iso}}\frac{\nabla T}{|\nabla T|},
\end{equation}
where we choose the factor $\phi_s=0.3$, the same as \citet{El-Badry2019MNRAS.490.1961E}.
The conductive heat flux is then determined by
\begin{equation}
    \frac{1}{\boldsymbol{Q}}=\frac{1}{\boldsymbol{Q_0}}+\frac{1}{\boldsymbol{Q}_\mathrm{sat}},
\end{equation}
i.e.,
\begin{equation}
    \boldsymbol{Q}=-\kappa\nabla T\frac{1}{1+\frac{\kappa|\nabla T|}{\frac{3}{2}Pc_{s,\,\mathrm{iso}}}}.
\end{equation}
In addition, we impose a ceiling such that $\kappa < 10^{11}\mathrm{\, erg\,cm^{-1}\,K^{-1}\,s^{-1}}$. We note that the ceiling on the thermal conductivity may suppress the effects of thermal conduction at the early stage, where the conductivity of the hot gas can be higher. We tested the effects of the ceiling and found that the effects of conduction can be slightly stronger by $\sim 10\%$ but the results are similar. We explicitly add the finite differencing conduction term into the energy flux, instead of operator splitting. Though for diffusive terms like thermal conduction, explicit time integrators typically are subject to a more restrictive stability limit $\Delta t \propto \Delta x^2$, the timestep here is not changed due to the saturation of the heat flux. 

In the pre-simulations of thermal instabilities, we use a constant conductivity $\kappa=\kappa_0=10^7\mathrm{\, erg\,cm^{-1}\,K^{-1}\,s^{-1}},$ to resolve conduction from even at low resolutions in which the gas pressure achieves the saturation pressure with the front between the CNM and WNM experience no net cooling.

\subsection{Radiative Cooling}
The net cooling rate takes the form of 
\begin{equation}
    \rho\mathcal{L}=n(n\Lambda(T)-\Gamma),
    \label{eq:cooling_func}
\end{equation}
where $n=\rho/(\mu m_\mathrm{H})$ the number density, $\Lambda(T)$ the cooling function, and $\Gamma$ the heating rate. For temperature below $10^{4.2}\,\mathrm{K}$, we use the formula from \citet{Koyama2002ApJ...564L..97K} considering atomic line cooling,
\begin{multline}
    \Lambda(T)=2\times10^{-19}\exp\left(\frac{-114800}{T+1000}\right)\\
    +2.8\times10^{-28}\sqrt{T}\exp\left(\frac{-92}{T}\right)\,\mathrm{erg\,s^{-1}\,cm^3}.
\end{multline}
For temperature between values of $10^{4.2}<T<10^{8.15}\,\mathrm{K}$, we use a linear interpolation of the tabulated cooling rate in Table 2 of~\citet{Schure2009A&A...508..751S} from highly-ionized ions assuming collisional ionization equilibrium for solar metallicity. For temperatures above $10^{8.15}\,\mathrm{K}$, we use a fit from~\citep{Schneider2018ApJ...860..135S},
\begin{equation}
    \Lambda(T)=10^{0.45\log{T}-26.065}\,\mathrm{erg\,s^{-1}cm^3},\,T>10^{8.15},
\end{equation}
to model the cooling dominated by bremsstrahlung. The heating rate is set as a constant
\begin{equation}
    \Gamma=\Gamma_0=5.0\times10^{-26}\mathrm{erg\,s^{-1}},
\end{equation}
for most of the simulations to represent the photoelectric effect of FUV radiation onto small grains. 

Accurate integration of the cooling term generally requires a time step much smaller than the hydrodynamic time step, imposing significant restrictions on the computational speed. Thus we use subcycling to compute the cooling rate if the cooling time step is smaller than the hydrodynamic time step. We tested our implementation of radiative cooling and thermal conduction by idealized simulations of thermal instabilities.

\subsection{Floors and Corrections}
In \athenak{} we evolve the conservative variables (mass, momentum, and total energy densities) using a high-order finite volume method including conductive fluxes while treating the cooling source terms with operator splitting. However, the presence of strong shocks, rarefaction, and cooling over a wide dynamic range can cause the primitive variables (density and pressure) to become unphysical and take negative values. To ensure the stability of the numerical results, we utilize the first-order flux correction (FOFC) algorithm as described by \citet{Lemaster&Stone2009ApJ...691.1092L}. In this algorithm, we first identify the problematic cells in which any of the density, temperature, or pressure is smaller than its floor after the full hydro update but before source terms. Then we roll back, replace the higher-order numerical fluxes at the interface of the problematic cells with more diffusive spatially first-order fluxes computed using the Local Lax Friedrichs (LLF) solver, and update the cells. In practice, this correction is required in less than $0.01\%$ of the cells ($\lesssim10^5$ every timestep in $2048^3\approx 10^{10}$ cells typically and $5\times10^5$ cells at maximum), and it does not affect the properties of the supernova remnants in normal cells.

Moreover, we apply floors when calculating the primitive variables from the conserved variables if the cell is still problematic after FOFC. We apply a density floor of $n_\mathrm{floor}=10^{-4}\,\mathrm{cm^{-3}}$, a temperature floor of $T_\mathrm{floor}\approx5\,\mathrm{K}$, and a pressure floor of $P_\mathrm{floor}/k_\mathrm{B}\approx 1\,\mathrm{K\,cm^{-3}}$. The floors are applied to $\lesssim 10^3$ cells in the grid of $2048^3$ cells.

\begin{deluxetable*}{lrccrccccrc}[htb]
    \tablenum{2}
    \tablecaption{List of simulations and their parameters. \label{tab:models}}
    \tablewidth{0pt}
    \tablehead{
    \colhead{Model} & \colhead{Mean} & \colhead{Cooling} &\colhead{Heating} & \colhead{Initial} & \colhead{Volume} & \colhead{Mass} & \colhead{Thermal} & \colhead{Grid} & \colhead{Resolution} & \colhead{Runtime} 
    \\[-0.2cm]
    \colhead{Label} & \colhead{Density} & \colhead{Function} & \colhead{[$\mathrm{erg\,s^{-1}}$]} & \colhead{Turbulence} & \colhead{Fraction} & \colhead{Fraction} & \colhead{Cond.} & \colhead{Size} & \colhead{[pc]}& \colhead{[kyr]} 
    }
    \startdata
    TN-512 & $10\,\mathrm{cm^{-3}}$ &  Yes & $5\times10^{-26}$ & 2-4 pc               & 0.11       & 0.92       & No         & $512^3$    & $0.125$ & 200\\
    TY-512 & $10\,\mathrm{cm^{-3}}$ &  Yes &  $5\times10^{-26}$ & 2-4 pc               & 0.11       & 0.92       & Yes        & $512^3$    & $0.125$ & 200\\
    TN-1024 & $10\,\mathrm{cm^{-3}}$ &  Yes &  $5\times10^{-26}$ & 2-4 pc               & 0.11       & 0.92       & No         & $1024^3$    & $0.0625$ & 200\\
    TY-1024 & $10\,\mathrm{cm^{-3}}$ &  Yes &  $5\times10^{-26}$ & 2-4 pc               & 0.11       & 0.92       & Yes        & $1024^3$    & $0.0625$ & 200\\
    TN-2048 & $10\,\mathrm{cm^{-3}}$ &  Yes &  $5\times10^{-26}$ & 2-4 pc               & 0.11       & 0.92       & No         & $2048^3$    & $0.03125$ & 200\\
    TY-2048 & $10\,\mathrm{cm^{-3}}$ &  Yes &  $5\times10^{-26}$ & 2-4 pc               & 0.11       & 0.92       & Yes        & $2048^3$    & $0.03125$ & 200\\
    TN-2048-X\tablenotemark{a} & $10\,\mathrm{cm^{-3}}$ &  Yes &  $5\times10^{-26}$ & 2-4 pc               & 0.11       & 0.92       & No         & $2048^3$    & $0.015625$ & 10\\
    TY-2048-X\tablenotemark{a} & $10\,\mathrm{cm^{-3}}$ &  Yes &  $5\times10^{-26}$ & 2-4 pc               & 0.11       & 0.92       & Yes        & $2048^3$    & $0.015625$ & 10\\
    \hline
    TN-w16 & $10\,\mathrm{cm^{-3}}$ &  Yes &  $5\times10^{-26}$ & 16-32 pc             & 0.11       & 0.92       & No         & $512^3$    & $0.125$ & 200\\
    TY-w16 & $10\,\mathrm{cm^{-3}}$ &  Yes &  $5\times10^{-26}$ & 16-32 pc             & 0.11       & 0.92       & Yes        & $512^3$    & $0.125$ & 200\\
    TN-h10 & $10\,\mathrm{cm^{-3}}$ &  Yes &  $1\times10^{-25}$ & 2-4 pc               & 0.05       & 0.84       & No         & $512^3$    & $0.125$ & 200\\
    TY-h10 & $10\,\mathrm{cm^{-3}}$ &  Yes &  $1\times10^{-25}$ & 2-4 pc               & 0.05       & 0.84       & Yes        & $512^3$    & $0.125$ & 200\\
    \hline
    TN-adb & $10\,\mathrm{cm^{-3}}$ &  No &  $5\times10^{-26}$ & 2-4 pc               & 0.11       & 0.92       & No         & $512^3$    & $0.125$ & 200\\
    TY-adb & $10\,\mathrm{cm^{-3}}$ &  No &   $5\times10^{-26}$ & 2-4 pc               & 0.11       & 0.92       & Yes        & $512^3$    & $0.125$ & 200\\
    \hline
    UN-1024 & $0.86\,\mathrm{cm^{-3}}$ &  Yes &   $5\times10^{-26}$ &  {---}             & 0.0       & 0.0       & No         & $1024^3$    & $0.125$ & 200\\
    \enddata
    \tablecomments{Some of the models and their parameters in this work. For the Models TN-512 and TY-512, we run them with 10 different initial seeds for initial perturbations.}
    \tablenotetext{a}{For these two runs, due to the limit of computational resources, we first increase the resolution of pre-simulation for model TN-512 and TY-512 from $512^3$ cells to $2048^3$ cells and rerun it for extra several Myrs to better resolve the boundary between the CNM and WNM. Then we cut the domain to $[-16,16]^3\,\mathrm{pc}$ and double the resolution to $2048^3$ to start the SNR simulation.}
\end{deluxetable*}

\begin{figure*}[htb]
    \centering
    \includegraphics[width=0.49\linewidth]{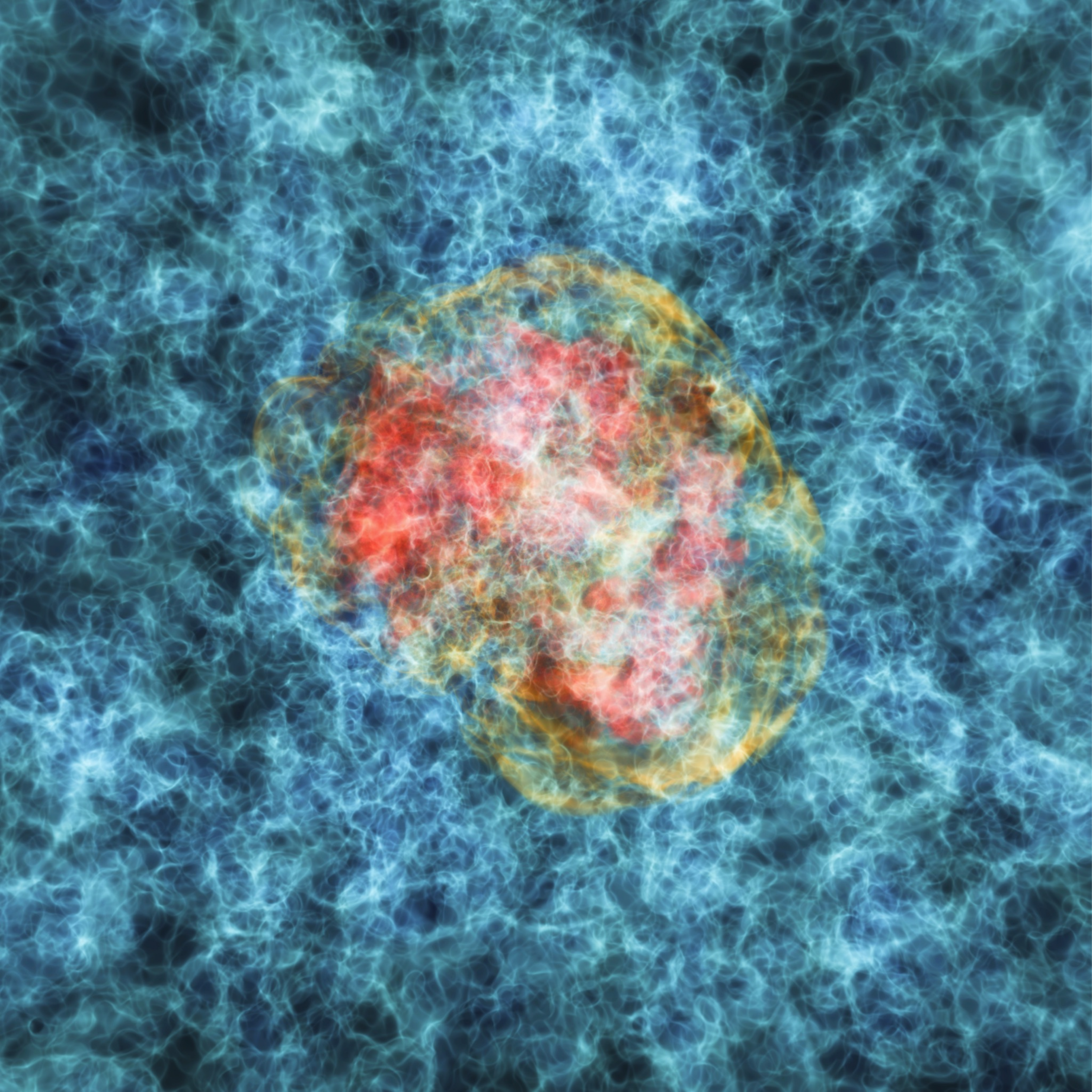}
    \includegraphics[width=0.49\linewidth]{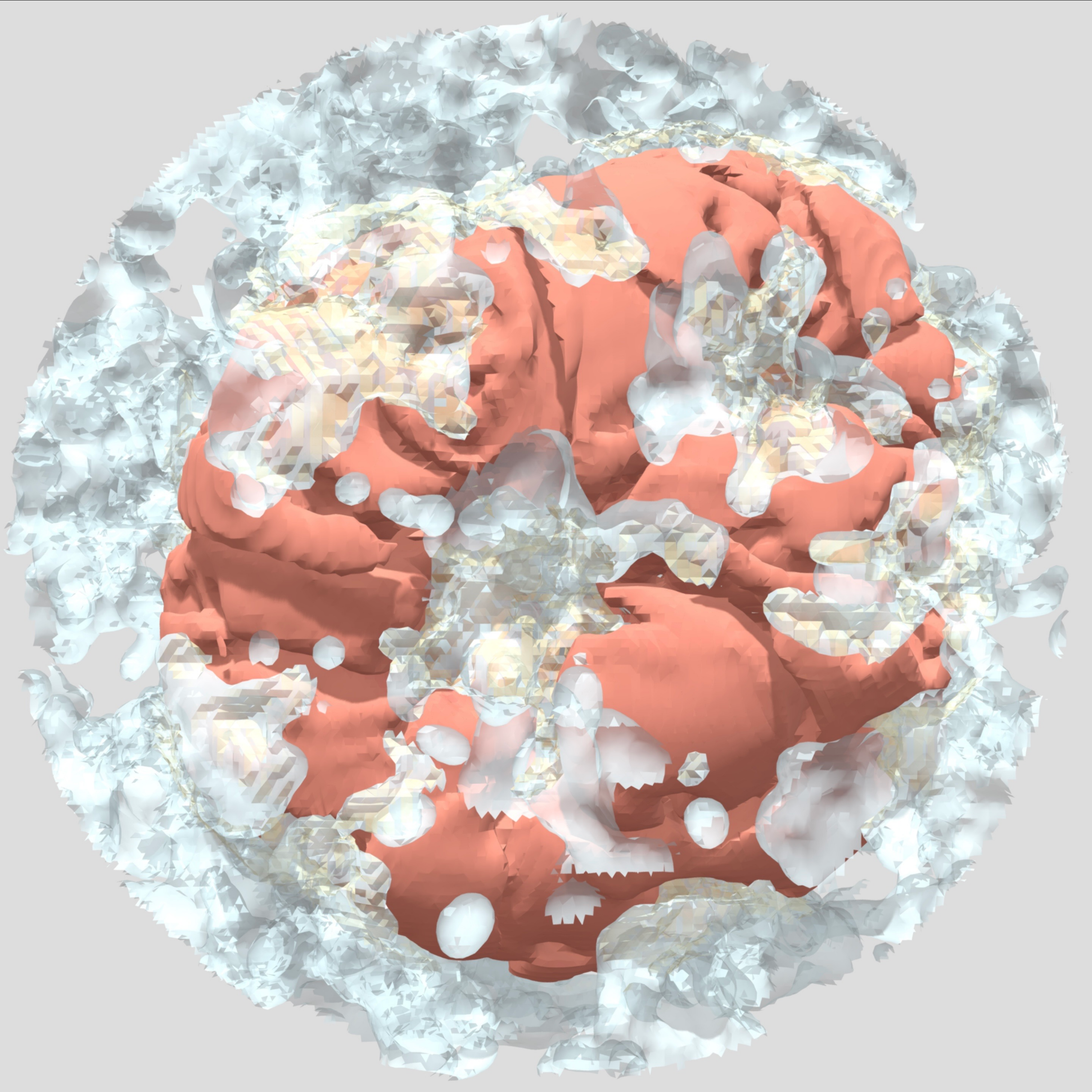}
    \caption{Left: 3D rendering of supernova remnants in the cloudy interstellar medium for model TN-2048. The rendering is made to highlight three ``layers'' of temperature: $T=5050$ (surface of CNM), $5\times10^6$ (shock front), and $3\times10^7 \,\mathrm{K}$ (hot gas interior) in blue, orange, and red, respectively. Right: Extraction of the surface of supernova remnants ($T=5\times10^6\,\mathrm{K}$, red) and the surface of the CNM ($T=5050\,\mathrm{K}$, light blue) within a sphere of $r=20\,\mathrm{pc}$. (The 3D model is available at \url{https://sketchfab.com/3d-models/snr512k0-t5e3k-t5e6k-4d3d1d52b1de490eb4f3af2dd0373eaf})}
    \label{fig:render}
\end{figure*}

\begin{figure*}[htb]
    \centering
    \includegraphics[width=\linewidth]{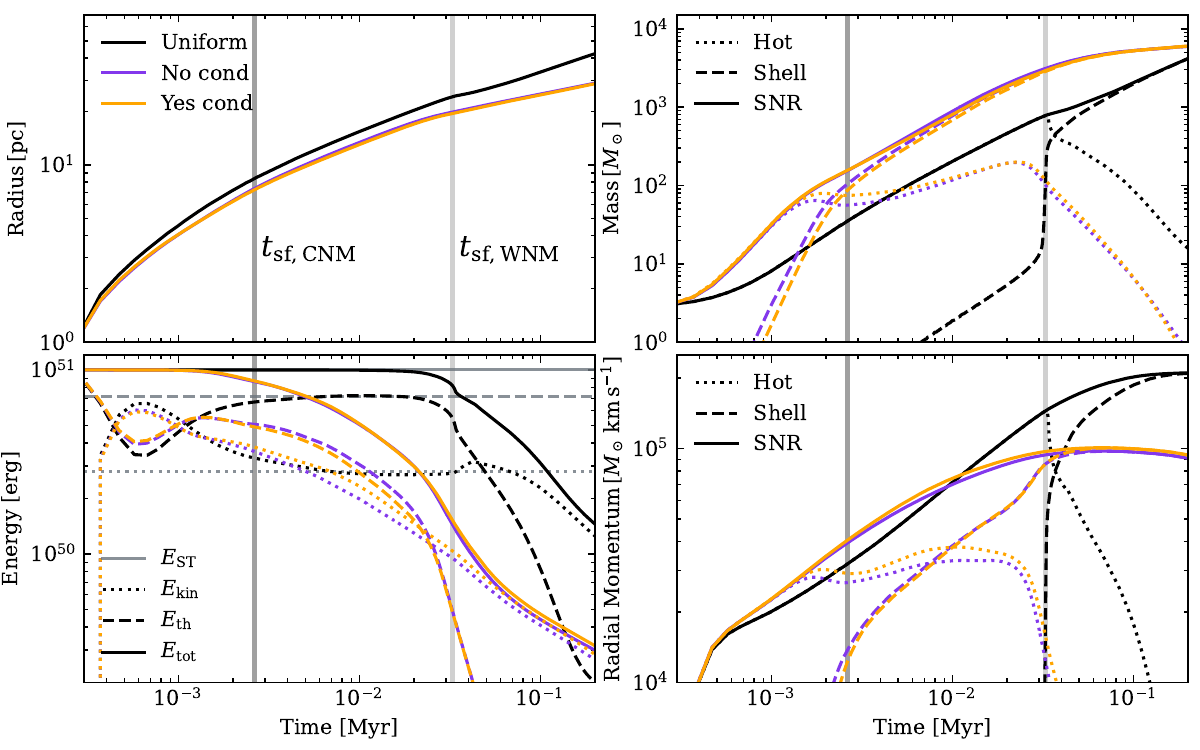}
    \caption{Evolution of radius, mass, energy, and radial momentum of supernova remnants in the cloudy inhomogeneous background without (purple) and with (orange) thermal conduction. The evolution in the uniform medium is shown by black lines. The vertical light and dark grey lines in the background correspond to shell formation time for a uniform background with the density of the WNM and CNM, respectively. The evolution in the uniform medium essentially follows the analytical theory. The cloudy background significantly increases the mass of the shell and reduces the energy more rapidly with less final momentum output. Thermal conduction makes little difference in the evolution of these global properties. Despite the considerable difference, the scaling of radius and mass between $t_\mathrm{sf,CNM}$ and $t_\mathrm{sf,WNM}$ is still similar to the analytical solutions.}
    \label{fig:evo_fidu}
\end{figure*}

\begin{figure*}[htb]
    \centering
    \includegraphics[width=\linewidth]{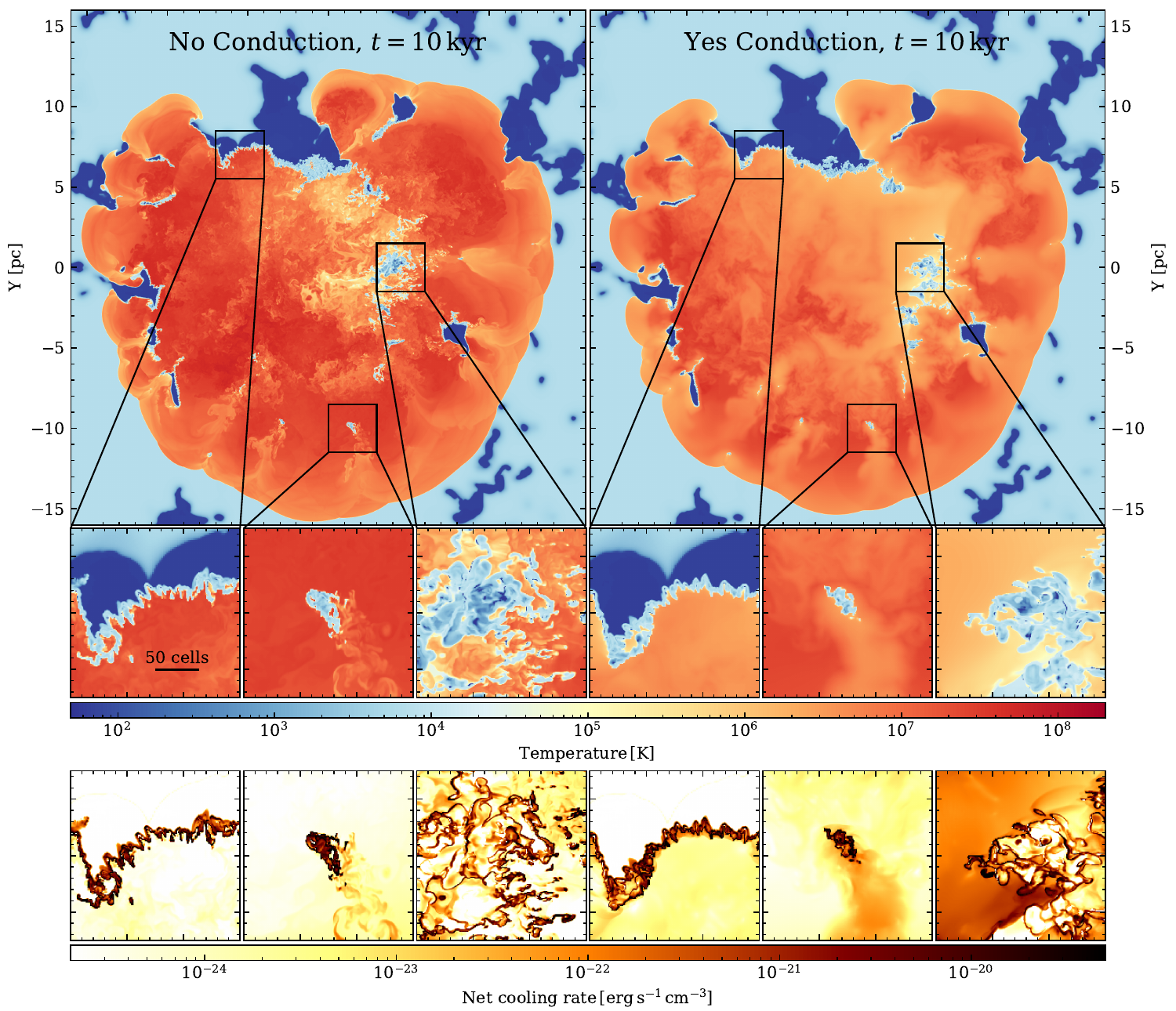}
    \caption{Top: slices of temperature through the $z=0$ plane of the SNRs at 10 kyr ($t/t_\mathrm{sf, WNM}\approx0.3$, $t/t_\mathrm{sf, WNM}\approx4$) without (left) and with (right) thermal conduction. Middle: zoom-ins of selected regions showing (from left to right) nonlinear thin-shell instability, single shock-cloud interaction, and turbulent mixing. Bottom: net cooling rate of the same selected regions.}
    \label{fig:slice}
\end{figure*}

\begin{figure*}[htb]
    \centering
    \includegraphics[width=\linewidth]{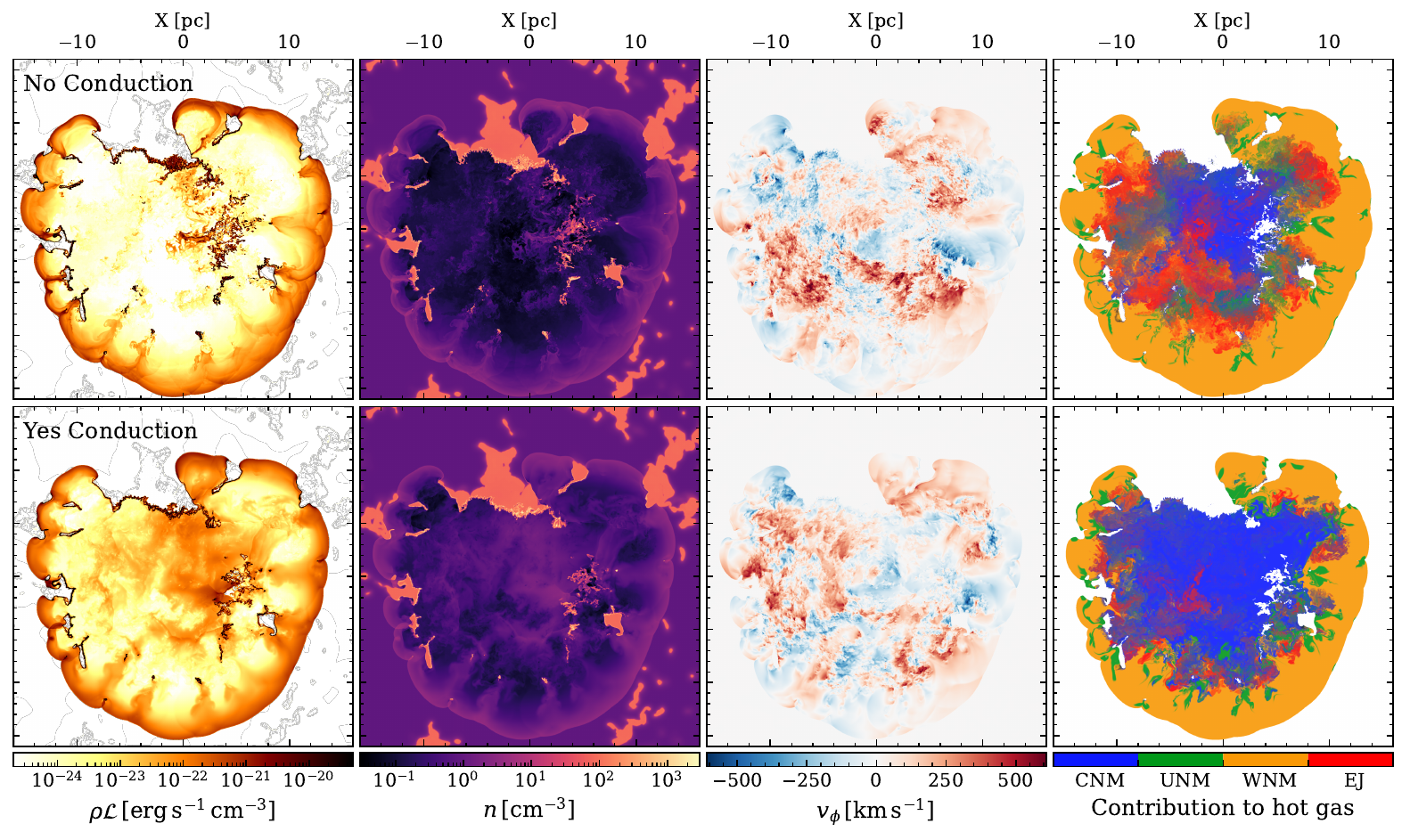}
    \caption{Slices through the $z=0$ plane of (from left to right) cooling rate, density, tangential velocity, $v_\phi$, and contribution to the hot gas from the four passive scalars without (upper) and with (lower) thermal conduction. Contribution to hot gas is calculated from the four passive scalars weighted by the mass fraction. The shock-cloud interaction is well-resolved with a resolution of $\Delta x= 1/64\,\mathrm{pc}$. Cooling is most efficient in the shell surrounding the cold clouds. Dense cool shells form around the cold clouds. The remnant is altered with a considerable tangential velocity, $v_\phi$. Thermal conduction smooths the interior of the hot gas considerably and slightly increases the efficiency of cooling in the hot gas regions.}
    \label{fig:slice_all}
\end{figure*}

\begin{figure*}[htb]
    \centering
    \includegraphics[width=\linewidth]{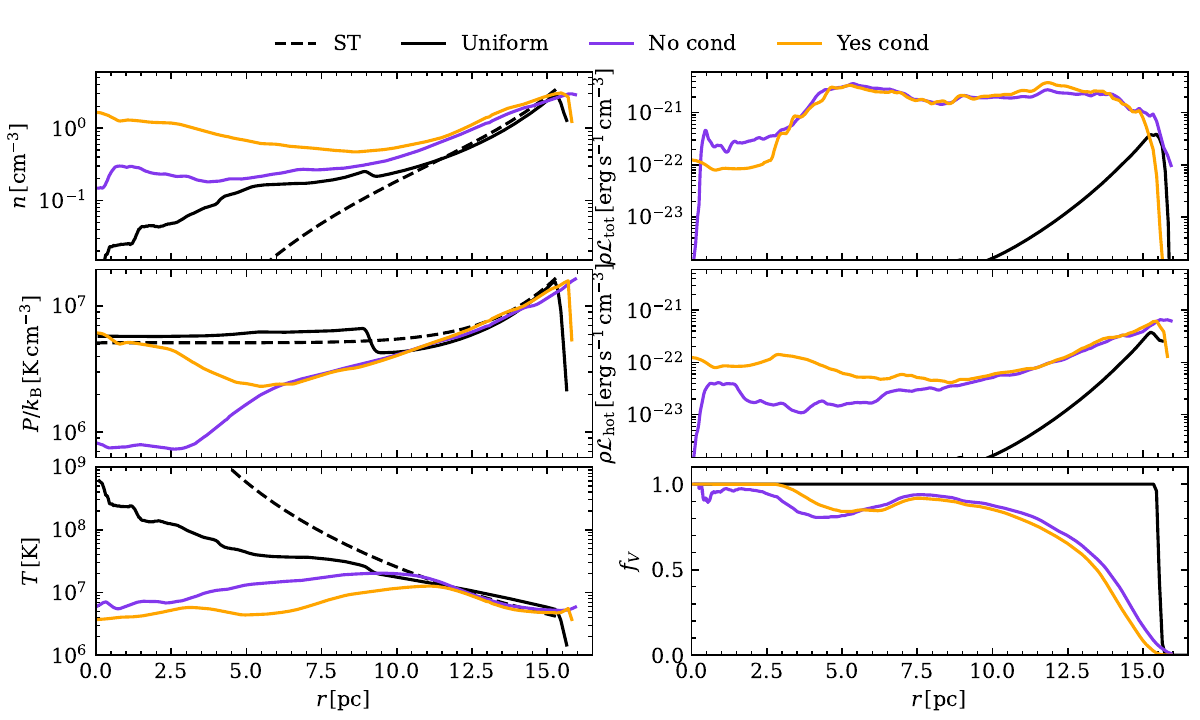}
    \caption{Radial profiles of hot gas density (top left), hot gas pressure (middle left), hot gas temperature (bottom left), total cooling rate (top right), hot gas cooling rate (middle right), and hot gas volume fraction (bottom right) at 10 kyr. For comparison, the radial profiles for the uniform case (UN-1024) and the ST solution are shown in black solid and dashed lines. The inner region is filled with hot gas with a relatively higher density and lower pressure and thus temperature compared with the ST solutions. Thermal conduction significantly increases the density and decreases the temperature, but the total cooling rate is still similar. The profile of the total cooling rate is flat within the SNR, except in the inner region where all the cold clouds are swept out. The volume fraction of hot gas decreases from about 10 pc.}
    \label{fig:radial}
\end{figure*}

\begin{figure*}[htb]
    \centering
    \includegraphics[width=\linewidth]{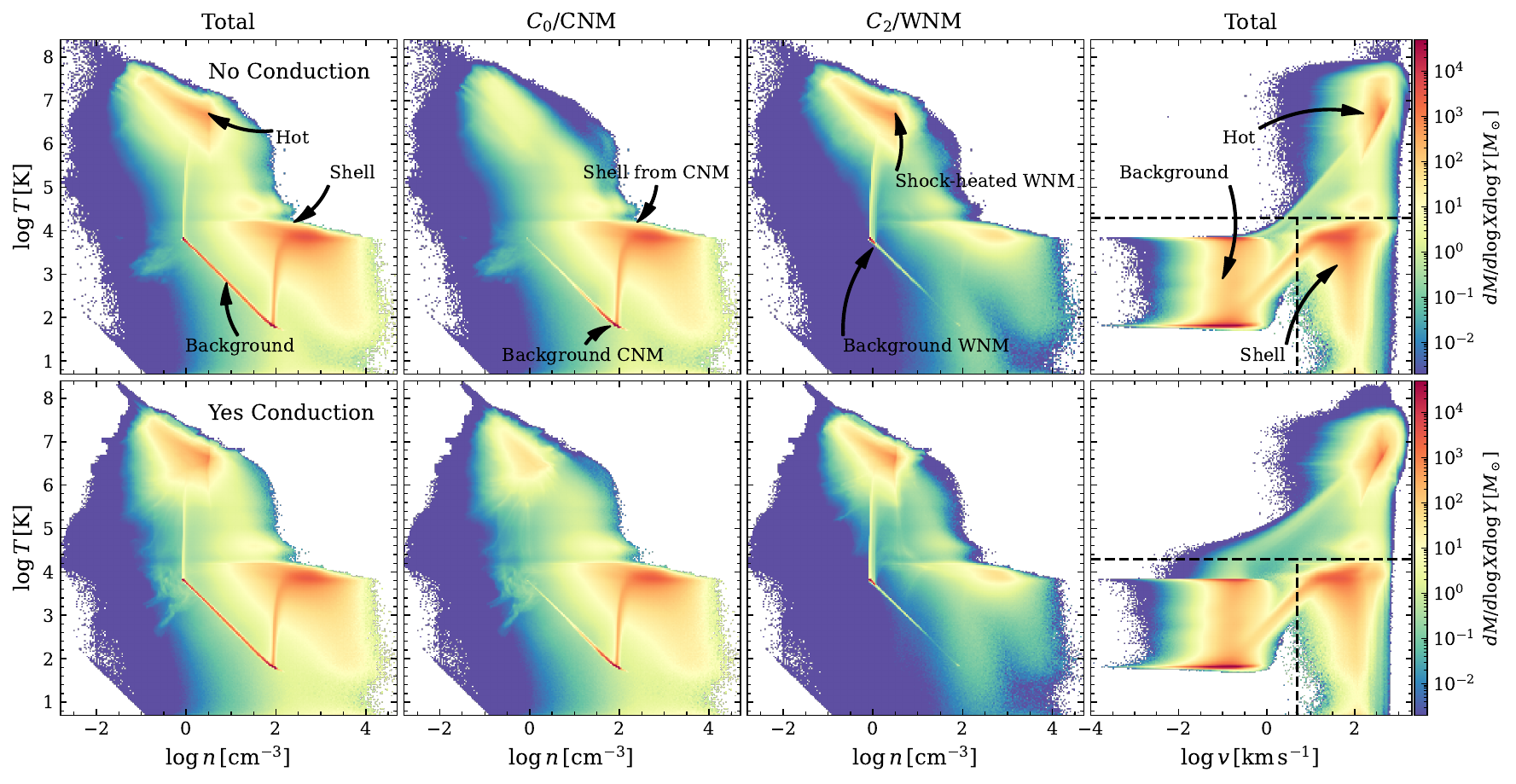}
    \caption{The $T-n$ joint PDF for all gas (first column), the gas that is initially CNM (species $C_0$) (second column), and the WNM (species $C_2$) (third column), and the $T-v$ joint PDFs (last column) of the gas at 10 kyr ($t/t_\mathrm{sf, WNM}\approx0.3$, $t/t_\mathrm{sf, CNM}\approx4$) without (upper panel) and with (lower panel) thermal conduction. Annotated are the main components of the ISM along with a significant fraction occupying intermediate regions. Dashes lines in the last column mark the boundary of $|v|=5\,\mathrm{km\,s^{-1}}$ and $T=5050\,\mathrm{K}$ used for definition of gas phases to distinguish the three components. Thermal conduction increases the contribution of CNM to hot gas.}
    \label{fig:dist}
\end{figure*}

\begin{figure*}[htb]
    \centering
    \includegraphics[width=0.49\linewidth]{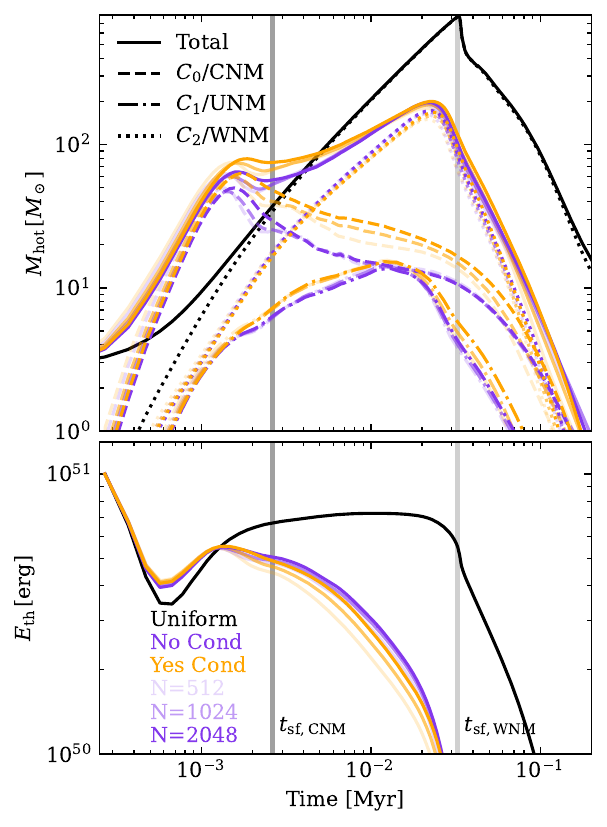}
    \includegraphics[width=0.49\linewidth]{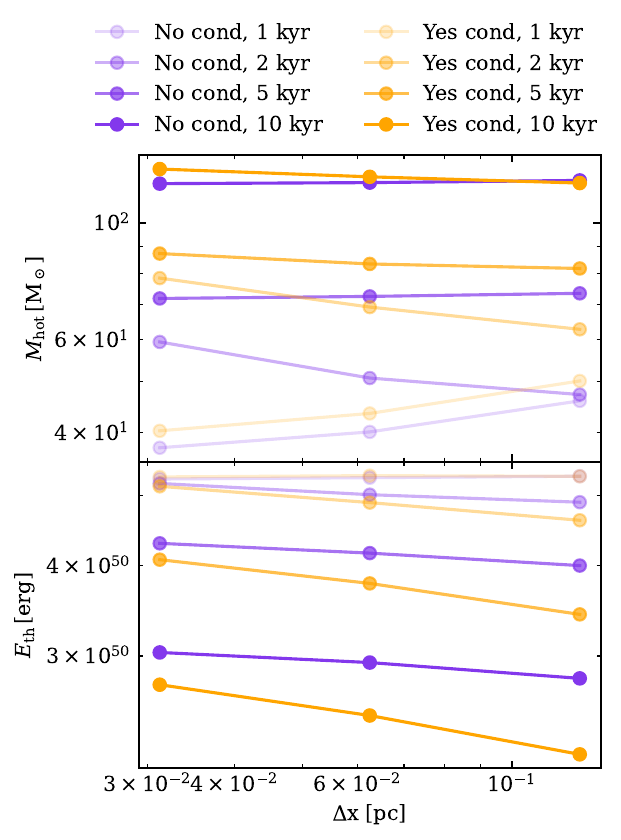}
    \caption{Left: Evolution of hot gas mass and contribution from different initial phases (top) and evolution of thermal energy of the supernova remnant for different resolutions. Right: Mass (top) and thermal energy (bottom) are functions of cell size at different times. The hot gas is first mainly from the shock-heated CNM, then mainly from the WNM. Due to the significant cooling, the hot gas mass is smaller than the uniform case. Higher resolution leads to a larger peak hot gas mass at the later time. Thermal conduction increases the hot gas mass by $\sim 30\%$ between $r_\mathrm{sf,CNM}$ and $r_\mathrm{sf,WNM}$. The is a change by a factor of $\sim20\%$ when doubling the resolution.}
    \label{fig:evo_dx}
\end{figure*}

\begin{figure*}[htb]
    \centering
    \includegraphics[width=\linewidth]{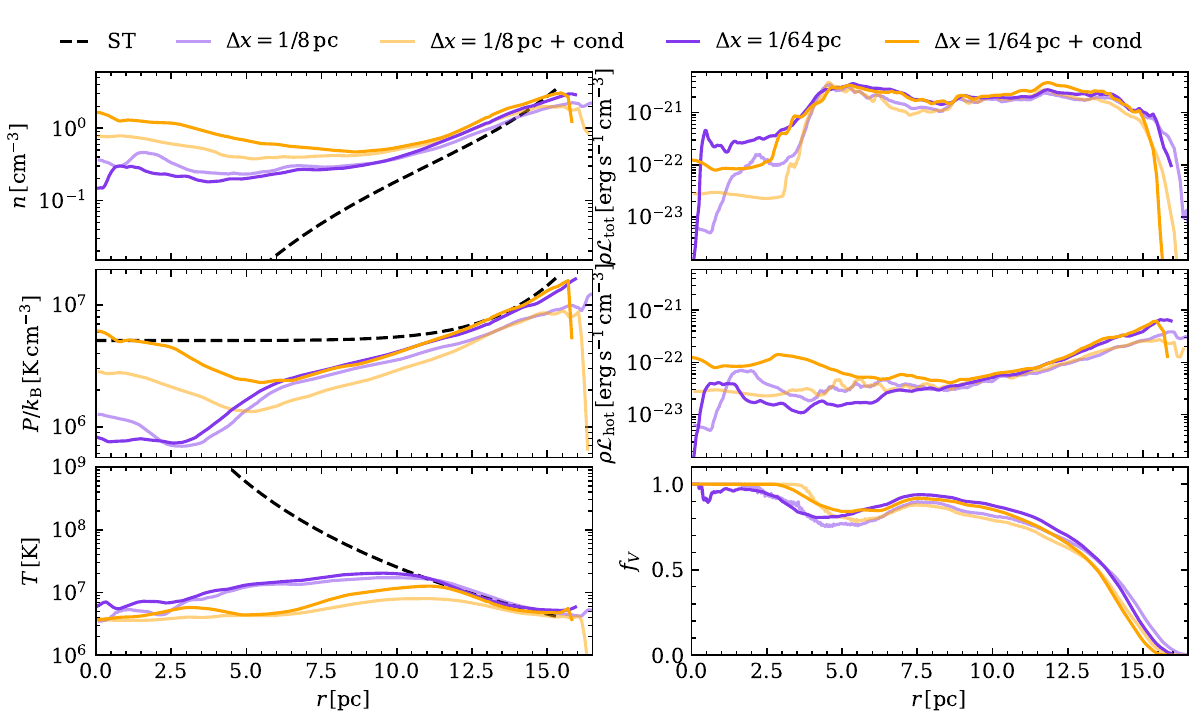}
    \caption{Similar to \figu\ref{fig:radial} but for two resolutions $\Delta x = 1/8\,\mathrm{pc}$ and $\Delta x = 1/64\,\mathrm{pc}$. The results are similar within a factor of $\sim2$, despite a factor of 8 difference in resolution. The models without conduction do not show a secular trend with resolution, while the models with conduction show systematically higher density, pressure, and cooling rates in most radii at higher resolution.}
    \label{fig:rad_dx}
\end{figure*}

\begin{figure*}[htb]
    \centering
    \includegraphics[width=0.9\linewidth]{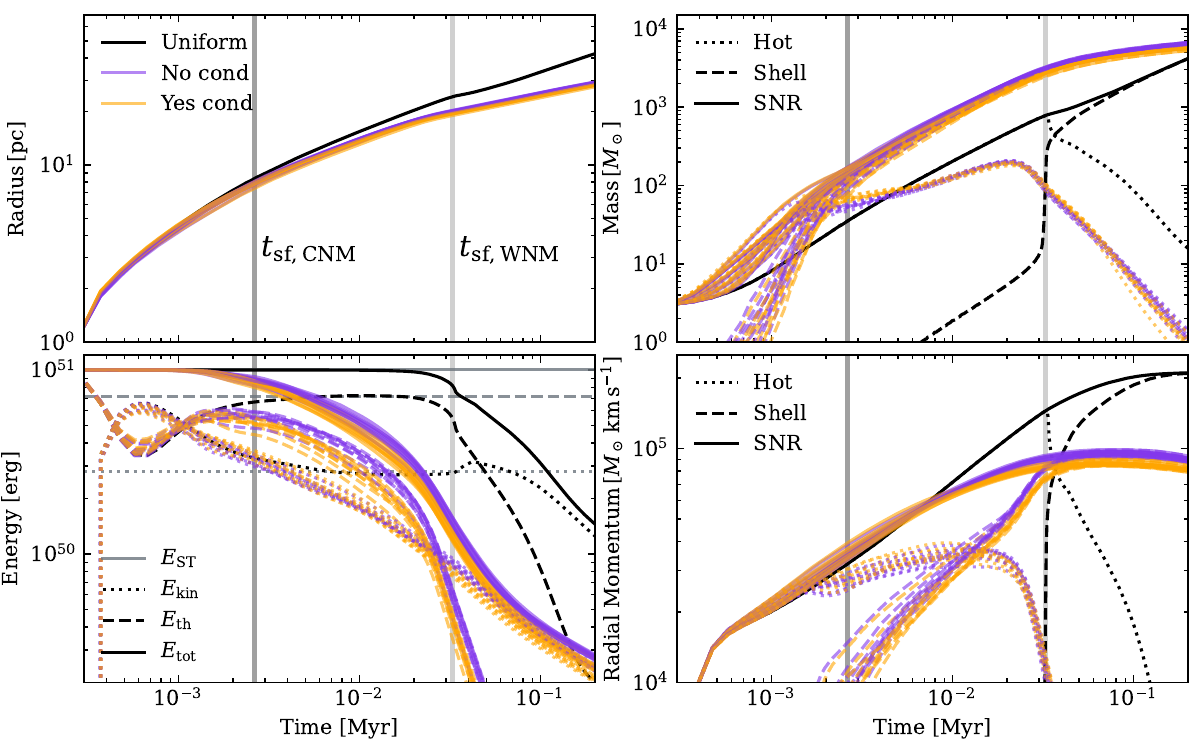}
    \includegraphics[width=0.9\linewidth]{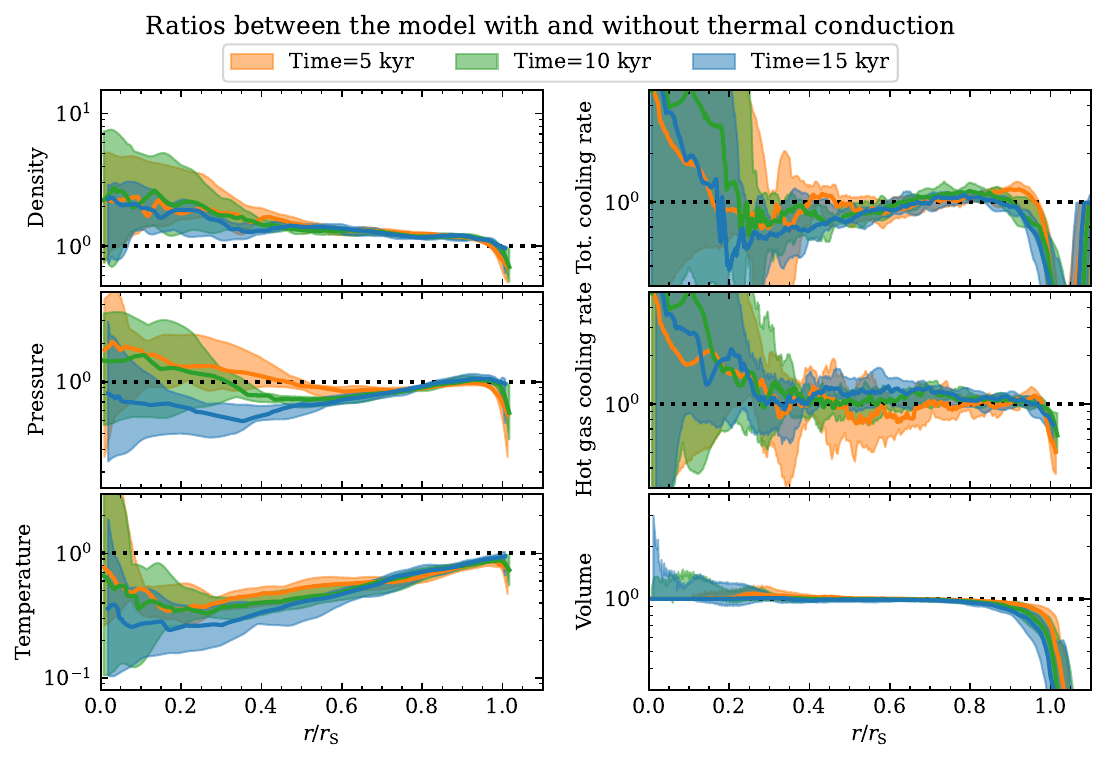}
    \caption{Top: similar to \figu\ref{fig:evo_fidu} but for 10 realizations with different initial perturbation seeds. Bottom: ratios between the model with and without thermal conduction for hot gas density (upper left), hot gas pressure (middle left) hot gas temperature (lower left), total cooling rate (upper right), hot gas cooling rate (middle right), and hot gas volume (lower right) as a function of radius at different times in the 10 realizations. The lines show the median value and the shaded regions show the range of all the 10 pairs of models. Despite a large variance by a factor of $\sim2-3$, the SNRs with thermal conduction show systematically higher density, lower pressure, and thus lower temperature. The total cooling rate is overall similar. The variance is slightly larger at the early stage in the central region due to the distribution of the cold clouds.}
    \label{fig:evo_seeds}
\end{figure*}

\begin{figure*}[htb]
    \centering
    \includegraphics[width=\linewidth]{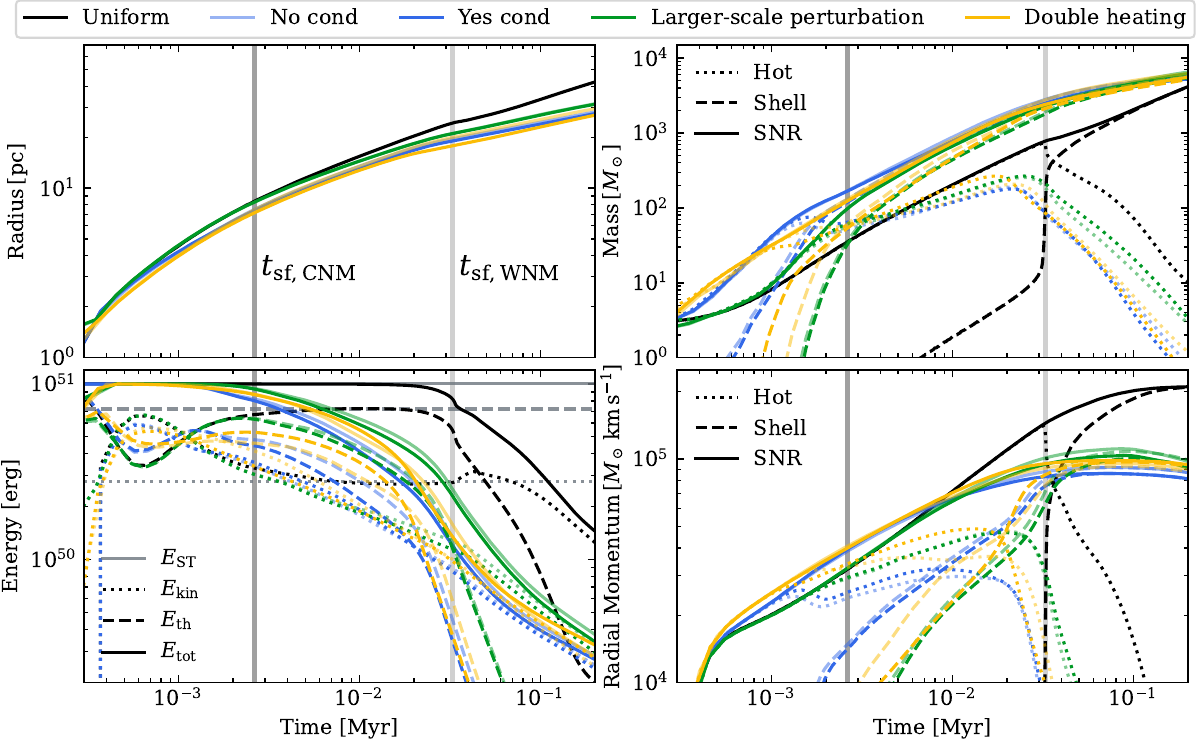}
    \caption{Similar to \figu\ref{fig:radial} but for different wavelengths of initial perturbation and heating rate (models T...-512, T...-w16, and T...-h10). The results are essentially similar with a difference of $\sim 10\%$ in radius, $\sim 50\%$ in mass, $\sim 50\%$ in energy, and $\sim 20\%$ in final momentum output.}
    \label{fig:evo_turb_hrate}
\end{figure*}

\begin{figure*}[htb]
    \centering
    \includegraphics[width=0.49\linewidth]{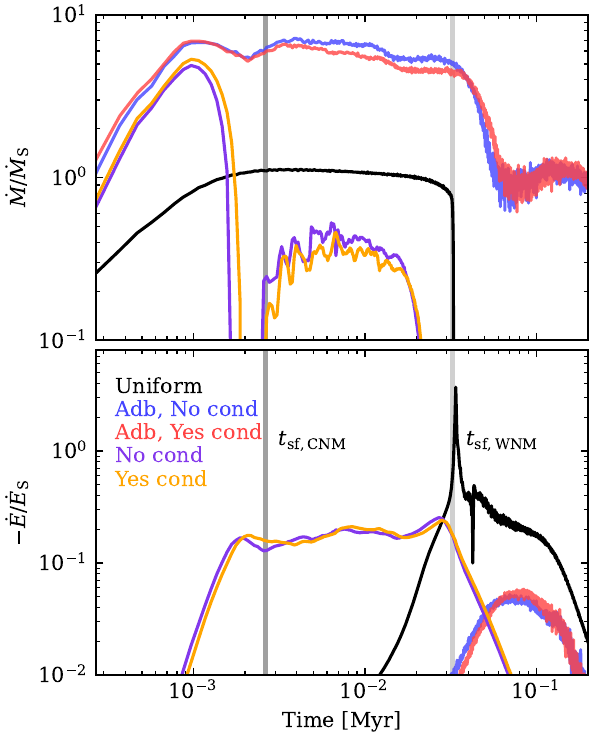}
    \includegraphics[width=0.49\linewidth]{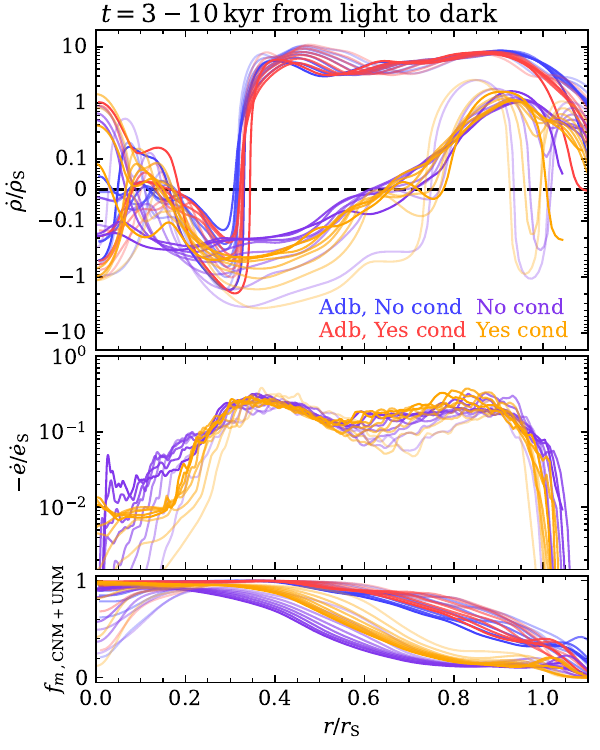}
    \caption{Left: history of hot gas mass growth rate $\dot{M}$ (top) and energy loss rate $-\dot{E}$ (bottom) normalized by $\dot{M}_\mathrm{S}$ and $\dot{E}_\mathrm{S}$ for the models T...-2048. Models T...-adb and UN-1024 are plotted for comparison. The mass growth and energy loss rate are relatively flat between $t_\mathrm{sf, CNM}$ and $t_\mathrm{sf, WNM}$ with $\dot{M}\sim 0.5 \dot{M}_\mathrm{S}$ and $\dot{E}\sim-0.2\dot{E}_\mathrm{S}$. Right: radial profiles of (from top to bottom) mass loading rate (\eq\ref{eq:mass_loading}), cooling rate, and mass fraction of the hot gas from CNM and UNM for the highest-resolution simulations (T...-2048-X) and adiabatic cases (T...-adb) for comparison. The lines are smoothed for clarity. Cooling significantly suppresses the mass loading and even turns mass loading into loss for the inner region. The mass loading rate  $\dot{\rho}\lesssim\dot{\rho}_\mathrm{S}$ in the outer region. The total cooling rate is spatially flat with $\dot{e}\sim-0.2\dot{e}_\mathrm{S}$.}
    \label{fig:evo_srcterms}
\end{figure*}

\begin{figure}[htb]
    \centering
    \includegraphics[width=\linewidth]{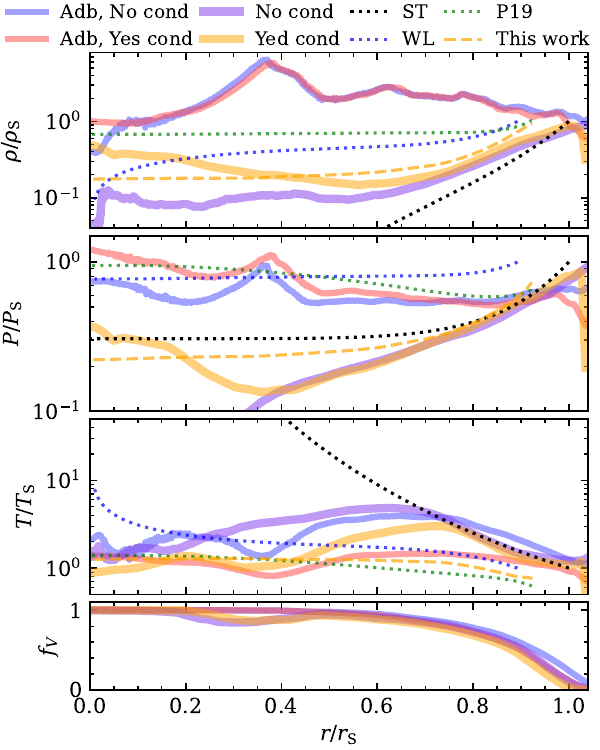}
    \caption{The radial profile of hot gas at 10 kyr in the 3D simulations overplotted with the ST solution, the WL solution~\citep{White&Long1991ApJ...373..543W} with an evaporating time of $\tau = 5$, the solution in \citet{Pittard2019MNRAS.488.3376P} with $f_\mathrm{ML}=10$ and the model proposed in this work which includes both mass loading and energy lost. The time-dependent model in this work can describe the evolution better, indicating the importance of cooling.} 
    \label{fig:rad_theory}
\end{figure}

\section{SN in two-phase medium} \label{sec:results} 

We explore the impact of embedded CNM on the time evolution and internal structure of SNRs before the shell formation of the WNM, $t_\mathrm{sf, WNM}$, delving into the roles and effects of two critical factors: turbulent mixing and thermal conduction. Furthermore, we examine how these effects vary with different resolutions, recognizing the importance of resolution in capturing the intricacies of the phenomena. The ability to capture and analyze those phenomena demonstrates the power and capability of the high-resolution approach utilized in our simulations. 

In terms of a single cloud crushing, \citet{Pittard&Parkin2016MNRAS.457.4470P} suggested that resolutions of 32-64 cells per cloud radius are the minimum necessary to capture the dominant dynamical processes in 3D simulations in local simulations of shock-cloud interaction. Our resolution ($\Delta x=1/64\,\mathrm{pc}$) is sufficient to achieve this requirement and capture the phenomena of ablation for clouds with size $\gtrsim 0.5\,\mathrm{pc}$ in the global simulation. As the majority of the clouds in our simulations are in this regime~(see \figu\ref{fig:cloud_dist}), we are capable of resolving the local shock-cloud interaction within global SNR evolution simulations. We note that we are not able to resolve the cooling length and conduction length, even in the foreseeable future. However, they may be less important since turbulent mixing is going to dominate in systems studied here, as suggested by~\citet{Fielding2020ApJ...894L..24F, Tan2021MNRAS.502.3179T}.

The models we run are summarized in \tab\ref{tab:models}. \figu\ref{fig:render} presents a representative three-dimensional rendering that displays the interaction between the supernova remnant and the cloudy two-phase medium. The supernova remnant is asymmetric due to the impediment of the cold clouds, showing sophisticated structures.

\subsection{Time evolution of global properties}
In \figu\ref{fig:evo_fidu}, we present the time evolution of several global properties of the SNRs, including radius, mass, energy, and radial momentum, for the models TN-2048 and TY-2048. We also include a model (UN-1024) of SNR expanding in a uniform background filled with WNM ($n_{\rm WNM}=0.86{\rm cm^{-3}}$) as a reference. The uniform medium case experiences the shell formation (marked by the sudden drop of thermal energy and the hot gas mass) at $t_{\rm sf,WNM}$ as analytically predicted. The energy ratio between thermal and kinetic also quickly converges to that expected by the analytic prediction ($E_{\rm th}:E_{\rm kin}=0.72:0.28$) during the ST stage after early relaxation from being fully thermal at injection.

We define the effective radius of the supernova remnant, $r_\mathrm{SNR}$, using the volume occupied by the ``SNR'' component as
\begin{equation}
    r_\mathrm{SNR}=\left(\frac{3}{4\pi}V_\mathrm{SNR}\right)^{1/3}.
\end{equation}
We note that despite the definition, the SNR is highly asymmetric. In comparison to the evolution in a uniform medium (UN-1024), the SNRs in a cloudy medium show smaller size and larger mass with decreasing total energy. The difference between the models with and without thermal conduction is negligible. The effective size of the SNR in a cloudy medium is only slightly smaller than that of the SNR in a uniform medium since the volume fraction of the CNM clouds is only $\sim10\%$. At $t_\mathrm{sf, WNM}$, the size of SNRs in a cloudy medium is $\sim 80-90\%$ of that in a uniform medium, though the expansion still follows a power law with a similar index with the ST stage, i.e., $r_{\rm SNR}\propto t^{2/5}$.

As more gas is available near the explosion point in the form of CNM clouds, the shock-heated hot gas mass increases more rapidly in the beginning. At around the shell formation time of the CNM ($t\gtrsim t_\mathrm{sf, CNM}=2.6\times10^{3}\,\mathrm{yr}$), the shock-heated CNM cools and the shell gas mass increases in turn. This leads to a gradual energy loss until the shell formation time for the WNM, $t_\mathrm{sf, WNM}=3.3\times10^4\,\mathrm{yr}$, at which the sudden energy loss occurs in the SNRs in both cloudy and uniform medium as forward shocks in the volume-filling WNM begin to cool. During the nominal ST stage before $t_{\rm sf,WNM}$, all swept-up mass becomes the hot gas in the uniform medium case, but in the cloudy medium case, the contribution from the swept-up CNM is accounted into the shell gas after $t_{\rm sf,CNM}$. As a result, the maximum hot gas mass at $t_{\rm sf,WNM}$ is $\sim 200 M_\odot$, a factor of $\sim 4$ smaller than the uniform case (UN-1024) in the cloudy medium case due to the energy loss even though the total swept-up mass at this time is about a factor of $\sim 4$ larger. It is about a factor of 2 smaller than the expected mass $M_{\rm sf}=410M_\odot$ for a uniform medium with the same mean density $n_0=10\,\mathrm{cm^{-3}}$ including both the WNM and the CNM.

The radial momentum at $t_\mathrm{sf, CNM}$ is larger than the uniform case (UN-1024) by a factor of $1.5$ owing to the increased SNR mass. However, the momentum at $t_\mathrm{sf, WNM}$ is $\sim 1\times10^5 M_\odot\mathrm{km/s}$, smaller than the uniform case by a factor of $2$ due to gradual energy loss in this stage. It is about $80-90\%$ of the expected momentum $p_{\rm sf}=1.1\times10^5M_\odot\mathrm{km/s}$ for a uniform medium with the same mean density $n_0=10\,\mathrm{cm^{-3}}$. There is no such a PDS stage because the early cooling makes the interior pressure already too low to allow much work on the shell.

Overall, a cloudy, two-phase background significantly increases the total swept-up mass of the SNR as more mass is available within a given volume in the form of cold gas. However, during the nominal energy conserving stage before $t_\mathrm{sf, WNM}$, it leads to a more rapid reduction in energy and a lack of radial momentum boost, resulting in less final momentum output. We note that our initial conditions are meant to be representative of the solar neighborhood. Generalization of the effect of cloudy background requires a wider parameter space study especially covering low-density and low-metallicity conditions.

\subsection{The internal structure of the SNR}
We here investigate the structure of the SNR using the highest-resolution simulations (TN-2048-X and TY-2048-X) conducted with a cell size of $\Delta x=1/64\,\mathrm{pc}\approx0.015\,\mathrm{pc}$. In \figus\ref{fig:slice} and \ref{fig:slice_all}, we present the intricate structure of the SNR and the interaction with the clumpy two-phase medium of the ambient environment at $t=10\,\mathrm{kyr}$. Subpanels in \figu\ref{fig:slice} from left to right highlight the shock propagation into a large cold cloud with corrugation due to non-linear thin shell instability, the ongoing evaporation of a single cloud, and the turbulent mixing region after the shocks run over large clouds completely. In all subpanels, cooling is mostly prominent in the interface between the hot gas and the cold clouds (see also the first column in \figu\ref{fig:slice_all}) owing to the high density of the dense shell and the high cooling coefficient at this intermediate temperature $10^4-10^5\,\mathrm{K}$. In addition, turbulent mixing processes generate cooling-efficient regions in wakes behind a single cloud and larger areas where multiple clouds interact. Thermal conduction further increases the density (second column in \figu\ref{fig:slice_all}) and decreases the temperature of the hot gas due to the evaporation of the cold clouds on top of the mixing due to hydrodynamical interactions. The enhancement of hot gas cooling is thus visible in larger volume (first column in \figu\ref{fig:slice_all}), while the total cooling is still dominated by the interface cooling such that the impact of thermal conduction on the overall time evolution of the SNR is relatively minor as shown in \figu\ref{fig:evo_fidu}. With conduction, the instabilities still arise, but the internal structure of the hot gas is smoothed. In addition, we note that the expansion of the SNR is altered with a considerable tangential velocity, $v_\phi=(x v_y - y v_x)/\sqrt{x^2+y^2}$ (third column in \figu\ref{fig:slice_all}). 

The inclusion of passive scalars in our simulations allows us to diagnose the contributions from different initial phases to the hot gas within the supernova remnant. As is shown in the last column of \figu\ref{fig:slice_all}, the central region of the remnant is primarily composed of material originating from the cold clouds (blue). Conversely, the outer region of the remnant is predominantly derived from the surrounding warm medium (orange). The initial hot ejecta (red) from the supernova explosion resides between these regions. The contribution from the unstable medium (green) can depict the striped gas from the cold clouds that are undergoing shock-cloud interaction as shown at the outer part of the remnant. Evidently, thermal conduction leads to an increase in the area where the contribution of cold gas dominates via the additional thermal evaporation of cold clouds, resulting in more rapid mixing of the hot ejecta.

\figu\ref{fig:radial} presents the angle-averaged radial profiles of the supernova remnant at 10 kyr. The internal density, pressure, and temperature profiles from the self-similar ST solution for the WNM are shown as dashed lines in the left column. Modulo the contribution from the ejecta (mainly near the center of the SNR), the ST solution is in good agreement with the uniform case. In a cloudy two-phase medium, the hot gas has a higher density, lower temperature, and lower pressure compared with the evolution in the uniform medium. Such density enhancement (mainly in the inner regions) is due to the mass added by the shock-cloud interactions. The inclusion of thermal conduction further increases the density as clouds experience conductive evaporation on top of the hydrodynamical interactions. As a result, the density and temperature profiles are flatter than the ST solution.

The density enhancement of the hot gas leads to a significant increase in the cooling rate from the hot gas (middle right), which was almost negligible in the uniform case except for the region right behind the shock. The total cooling rate (top right) is, however, dominated by the cooler gas (which we refer to as the shell gas) at the interface between the hot and cold gas (see \figus\ref{fig:slice} and \ref{fig:slice_all}) with little dependence on thermal conduction. The total cooling rate profile remains relatively flat throughout the SNR, except for the central region, where all the cold clouds are either eradicated or pushed away (see high volume fraction of the hot gas in the bottom right panel). The endpoint of the radial profile of the ST solution as well as the uniform case marks the predicted shock radius for the WNM, $r_\mathrm{S}\approx 15\,\mathrm{pc}$ at $t=10\,\mathrm{kyr}$. Even for the cloudy medium cases, the SNRs are predominantly occupied by the hot gas with a volume fraction of approximately 80-90\% until $0.8r_\mathrm{S}$, which decreases to zero at around $r_\mathrm{S}$. This decline suggests that the shocks in the direction dominated by the CNM have not reached $r_\mathrm{S}$. The half-volume radius is $\sim 0.9\,r_\mathrm{S}$, consistent with the value of the SNR radius, $r_\mathrm{SNR}$. 

\figu\ref{fig:dist} shows the $T-n$ joint PDF for all gas, the gas that is initially CNM (species $C_0$), and the WNM (species $C_2$), and the $T-v$ joint PDFs. The $T-n$ phase diagram comprises three segments annotated in the plot: the background equilibrium section under the initial conditions, the high-velocity, high-temperature, low-density gas section, and the high-velocity dense shell gas section. With the advantage of high resolution in our simulations, we are able to resolve regions of extremely dense gas within the supernova remnant, characterized by densities as high as $n \gtrsim 10^4\,\mathrm{cm^{-3}}$. Again, utilizing passive scalars enables us to trace the evolution of different gas phases and their contributions within the supernova remnant. The simulations indicate that both the CNM and WNM undergo isochoric shock heating. While the shell predominantly comes from the CNM, the high-temperature gas primarily originates from the shock-heated WNM at $\sim 10$ kyr. Thermal conduction significantly enhances the contribution of the CNM to the hot gas, though the source of hot gas is still dominated by the WNM at this time.

\subsection{Dependence of hot gas evolution on resolution}\label{subsec:convergence}

According to the criteria by \citet{Pittard2019MNRAS.488.3376P}, we reach the resolution required to resolve cloud crushing for most of the clouds in our simulations (see \figu\ref{fig:cloud_dist}). However, our resolution is insufficient to formally resolve the cooling length ($l_{\rm cool} = c_s t_{\rm cool} \approx 7\times10^{-5}\,\mathrm{pc}$ ) and the Field length ($l_{\rm F} \approx 3\times10^{-5}\,\mathrm{pc}$ ) at the cloud interface for a typical number density of $n=10^2\,\mathrm{cm^{-3}}$ and temperature of $T=2\times10^4\,\mathrm{K}$. Still, \citet{Fielding2020ApJ...894L..24F} and \citet{Tan2021MNRAS.502.3179T} showed that, even though the interface is not fully resolved (which results in broadening of the interface and pressure dips at the interface), the total amount of cooling and energy flux across the interface are well converged at lower resolutions. As a result, the numerical requirements to get mass flux and energy flux correct are likely not so stringent. 
Here, we are mainly interested in the dynamical impacts of the shock-cloud interaction on the SNR evolution. Therefore, we examine the resolution dependence using the evolution of mass and energy of SNRs as well as the internal structures.

In \figu\ref{fig:evo_dx} we show the time evolution of hot gas mass and thermal energy and its dependence on conduction and resolution for the models using the same initial clouds distribution, i.e., TN-512, TY-512, TN-1024, TY-1024, TN-2048, and TY-2048. On the left panel, we show the full time evolution, while we compare the hot gas mass and energy at selected times on the right panels. It is worth noting that although we try to make problems identical across the resolution, the warm-cold interface structure in the pre-simulations is not fully converged in the sense that the change of the resolution only depends on the numerical truncation errors. Because of this, we cannot claim any formal numerical convergence of the problem. Rather, we discuss the resolution dependence of physical properties and claim convergence in terms of the invariance of the qualitative behavior with resolution.

We first focus on the models without conduction (TN, purple colors). Compared with the uniform case, the hot gas mass is higher at the earlier stage due to the dominant contribution from the shocked CNM. The larger peak mass is achieved at later times at higher resolutions. At this stage, there is a noticeable difference in the hot gas mass with resolution (higher hot gas mass at higher resolution). As the shocked CNM begins to cool, the hot gas mass is in turn dominated by the shocked WNM. The contribution from UNM (the gas at the initial CNM-WNM interface) also becomes considerable. About $2t_\mathrm{sf, CNM}$ later, the contributions from all initial phases become resolution independent. At this stage, the contributions from the CNM and UNM are set by the cloud crushing and turbulent mixing due to the shock-cloud interaction. This process is likely converged in our simulations as we have high resolution to resolve most clouds and the mass and energy exchange rates in the turbulent mixing layer seem to converge. It is likely conspired by the compensation between the decrease in numerical dissipation and the increase in surface area at higher resolution \citep{Lancaster2024ApJ...970...18L}.

With conduction (TY; orange colors), the contribution of the CNM to the hot gas is increased significantly by $\sim 30\%$ compared to the TN modes during the time in between $t_\mathrm{sf, CNM}$ and $t_\mathrm{sf, WNM}$. This represents additional conductive evaporation of cold clouds embedded in the hot gas. The resolution dependence of the CNM contribution in the TY models is still visible in the hot gas mass (increasing $\sim 20\%$ when doubling the resolution). However, the contribution from the conductive evaporation to the total hot gas mass is still much smaller than the shocked WNM and is dynamically less important (see \figu\ref{fig:evo_fidu}). While unexplored in this paper, the CNM contribution to the hot gas is dominant near the center of SNR and likely affects the enhanced X-ray luminosity of the central region in the mixed-morphology SNRs \citep[e.g.,][]{Rho1994ApJ...430..757R, Vink2012A&ARv..20...49V}. 

In terms of total energy, there is a general trend of less energy loss at higher resolution, but the difference between two resolution runs gets smaller as the resolution gets higher. Also, the difference between with and without conduction is larger at lower resolutions. Because the conductive mass evaporation rate is determined by the \emph{net} energy flux from the hot gas (the conductive heat flux minus interface cooling), the higher cooling at lower resolution leads to the reduction in the net heat flux into the cold clouds, which is offset by lower mass evaporation as seen in the resolution dependence of the hot gas mass. 

\figu\ref{fig:rad_dx} plots the radial profiles for two resolutions $\Delta x=1/8\,\mathrm{pc}$ (T...-512) and $\Delta x=1/64\,\mathrm{pc}$ (T...-2048-X) at 10 kyr. Despite a factor of 8 in resolution and a slight difference in initial conditions due to pre-simulations, there is a factor of $\lesssim 2$ in the difference of radial profiles. The difference in the radial profiles is most prominent near the center of SNRs and with the models with thermal conduction. The TN models do not show secular trend with resolution, while the TY models show systematically higher density, pressure, and cooling rates in most radii at higher resolution. Again, the difference in the TY models is mainly caused by the additional cloud evaporation process, which is not fully resolved. We note here though that the current isotropic thermal conduction used in our paper (albeit the heat flux saturation) should be considered as a maximal conduction treatment, which can be suppressed in the presence of magnetic fields \citep{Orlando2008ApJ...678..274O}.

In summary, the resolution dependence of the SNR evolution in our simulations is qualitatively converged, especially in the models without conduction. For example, we do not expect a sudden change in the expansion law (e.g., from ST-like to \citet{McKee&Ostriker1977ApJ...218..148M}-like) as we further resolve the shock-cloud interaction. Our measurements of the global properties (e.g., the amount of hot gas, \figu\ref{fig:evo_dx}) and angle-averaged radial structure (\figu\ref{fig:rad_dx}) are robust and show small differences ($\lesssim 20\%$ when doubling the resolution). However, even higher resolutions and more careful treatments of thermal conduction are warranted to make quantitative predictions of observables, for example, X-ray properties of the mixed-morphology SNRs.

\subsection{Effects of different realizations and parameters}
So far, we have examined the SNR evolution in one realization of a background cloudy medium. The initial distribution of the cold clouds can significantly alter the evolution of the supernova remnant, especially at the early stage, when a single cloud or a few clouds play a dominant role. To gauge the level of this effect, here we run 10 models with different random seeds for the initial perturbation. 

\figu\ref{fig:evo_seeds} plots the time evolution of radius, mass, energy, and radial momentum of these 10 models. Overall there is a relative difference of $\sim10\%$ in radius, $\sim 50\%$ in mass, $\sim \%30$ in energy, and $\sim 20\%$ in the final radial momentum output. The early evolution before $t_\mathrm{sf,CNM}$ is almost identical between the TN and TY models. At the later stage, as the evolution is more and more controlled by the collective behaviors of all the clouds within the SNR, the systematic difference due to the presence of thermal conduction becomes prominent. Additional conductive evaporation leads to higher interior density and stronger cooling, giving rise to smaller mass, thermal energy, and radial momentum that are $\sim 80-90\%$ of those in the case without thermal conduction at $t_\mathrm{sf, WNM}$. 

The radial profiles of ratios for various properties between the model with and without thermal conduction for these 10 models are shown in \figu\ref{fig:evo_seeds}. Despite a large variance by a factor of 2 to 3 in radial structure due to different realizations, the systematic effects of thermal conduction still exist. Again, the SNRs with thermal conduction show systematically higher density, lower temperature, and slightly higher cooling rate. However, the difference between different realizations can dominate over thermal conduction in the inner region in certain extreme cases, owing to the randomness of the distribution of the cold clouds, as expected. We note that the resolution here is relatively low and thus the effects of thermal conduction and cooling may be slightly biased, as is shown in \figus\ref{fig:evo_dx} and \ref{fig:rad_dx} and discussed in \sect\ref{subsec:convergence}.

In addition, we also vary the heating rate and the wavelength of initial perturbation. The results are shown in \figu\ref{fig:evo_turb_hrate}. Varying the heating rate changes the saturation pressure and thus the density of the two phases with the volume and mass fraction adjusted accordingly in the pre-simulations. Larger-scale initial perturbation does not systematically change the background but behaves like a model with a different seed since the clouds are developed from the nonlinear thermal instabilities. These two parameters overall lead to a change of $\sim 10\%$ in radius, $\sim 50\%$ in mass, $\sim 50\%$ in energy, and $\sim 20\%$ in final momentum output, similar to the level of difference from different seeds.

\section{Implication for semi-analytical models}\label{sec:discussion}

Because the volume filling fraction of the cold, dense clouds in the ISM is typically small, the effect of clouds in SNR evolution can be treated as ``impurities'' by considering additional source and sink terms. Early analytic theories include the case where the global expansion of SNRs is modified by cloud evaporation \citep{McKee&Ostriker1977ApJ...218..148M} or the case where the self-similar internal profiles are modified while the expansion still follows the ST solution \citep{White&Long1991ApJ...373..543W}. As a general framework, one can solve a set of 1D spherical equations for the volume-filling component by incorporating the effects originated by shock-cloud interaction in terms of mass, momentum, and energy source and sink terms \citep[e.g.,][]{Dyson2002A&A...390.1063D, Pittard2003A&A...401.1027P, Pittard2019MNRAS.488.3376P}. At the same time, one can build a model for the cloud evolution (or population of clouds) to complete the description of the two-way interactions \citep[e.g.,][]{Cowie&McKee1977ApJ...211..135C, Cowie1981ApJ...247..908C, Fielding2022ApJ...924...82F}.

In this section, we make a minimal modification to the source/sink terms to find the simplest model that can describe the volume-filling component of our simulation results. We introduce two additional terms in the 1D spherical evolution equations: mass source and energy sink. The equations in this case are
\begin{align}
    \frac{\partial \rho}{\partial t} + \frac{1}{r^2}\frac{\partial}{\partial r}(r^2 v\rho)&=\dot{\rho},\\
    \frac{\partial \rho v}{\partial t} + \frac{1}{r^2}\frac{\partial}{\partial r}(r^2\rho v^2) + \frac{\partial P}{\partial r}&=0,\\
    \frac{\partial \mathcal{E}}{\partial t} + \frac{1}{r^2}\frac{\partial}{\partial r}\left[r^2 v\left(\mathcal{E}+P\right)\right]&=\dot{e},
\end{align}
where $\rho$, $v$, $P$, and $\mathcal{E}=e+\rho v^2/2=P/(\gamma-1)+\rho v^2/2$ are the ISM density, velocity, pressure, and total energy density. The additional terms, $\dot{\rho}$ and $\dot{e}$, are the source or sink terms for mass and energy. Although we do not attempt to construct analytic models for specific processes responsible for these terms, the terms are representative of mass source through cloud crushing and evaporation and energy loss by cooling at the mixing layers, which are directly calibrated to the simulation.

To understand our simulation results in the context of the 1D impurity models, we first measure the corresponding source and sink terms. In the left column of \figu\ref{fig:evo_srcterms}, we show the evolution of the integrated mass growth rate $\dot{M}=dM/dt$ and energy loss rate of the hot gas $\dot{E}=dE/dt$ by taking the derivative of the hot gas mass and supernova remnant gas energy history with regard to time and normalize them by $\dot{M}_\mathrm{S}$ and $\dot{E}_\mathrm{S}$. As a reference, we show a run in a uniform background (model UN-1024), in which there is no energy loss before $t_\mathrm{sf,WNM}$ and a sharp rise of $-\dot{E}$ at $t_\mathrm{sf,WNM}$. The hot gas mass growth rate is consistent with the swept-up mass during the ST stage and plunges suddenly at $t_\mathrm{sf,WNM}$ as the swept-up mass after this time is accumulated as the shell gas. In the fiducial models (T...-2048), there is an early rise and fall of $\dot{M}$ before $t_\mathrm{sf, CNM}$ due to the cold clouds. Then between $t_\mathrm{sf, CNM}$ and $t_\mathrm{sf, WNM}$, there are nearly constant mass growth rate and energy loss rate. The mass growth rate $\dot{M}\sim 0.5 \dot{M}_\mathrm{S}$ is smaller than $\dot{M}_\mathrm{S}$ because of the smaller solid angle to which the shock propagates into the WNM. The mass loading by shock-cloud interaction is not dominant during this time. Any shocks propagating into the CNM cool quickly, leading to continuous energy loss with $\dot{E}\sim -0.2\dot{E}_\mathrm{S}$. We also plot results from adiabatic runs (TN-adb and TY-adb) as a reference for ``maximal'' mass loading $\dot{M}\sim 5 \dot{M}_\mathrm{S}$ without energy loss. We note that there is a drop in mass loading and a rise in energy loss at a later time due to the fact that the SNR becomes bigger than the box size, but the evolution at this time is beyond the scope of this work.

We now measure the local mass loading rate. However, measuring the local mass loading rate directly from the individual shock-cloud interactions in the current simulations is difficult without additional analysis tools (e.g., tracer particles). Instead, we measure the mass loading rate by using the time-dependent angle-averaged density profiles and taking numerical finite difference following
\begin{equation}
    \dot{\rho_\mathrm{i}}=\frac{1}{4\pi }\int_{\Omega_\mathrm{i}}\left[ \frac{\partial\rho_\mathrm{i}}{\partial t}+\frac{1}{r^2}\frac{\partial (r^2\rho_\mathrm{i} v_{r,\mathrm{i}})}{\partial r}\right]\dd \Omega,\label{eq:mass_loading}
\end{equation}
where the subscript $\mathrm{i}$ could be any species but here we focus on the hot gas. In the right panel of \figu\ref{fig:evo_srcterms} we plot the radial profiles of the mass loading rate of hot gas estimated from the highest-resolution simulations and adiabatic runs as comparison. Without cooling, mass loading shows a spatially flat profile and mainly happens in the middle and outer regions of the SNR since the clouds are destroyed in the central region. Cooling suppresses mass loading in the middle region and even turns mass loading into mass loss for the inner region. The magnitude of mass loading in the outer region is around $\dot{\rho}_\mathrm{hot}\lesssim \dot{\rho}_\mathrm{S}$. The total cooling rate profile $\dot{e}=-n^2\Lambda_\mathrm{tot}$ is spatially flat and scales with $\dot{e}_\mathrm{S}$ as $\dot{e}\sim-0.2\dot{e}_\mathrm{S}$. We note that the mass contribution from the cold clouds to the hot gas gradually decreases as the radius increases.

Motivated by the mass loading rate and cooling rate measured in \figu\ref{fig:evo_srcterms}, we construct a time-dependent model in which mass loading and energy loss are considered, i.e.,
\begin{equation}
    \dot{\rho}(r,t)=\mathcal{A}\dot{\rho}_\mathrm{S}k(r/r_\mathrm{S}),
\end{equation}
and
\begin{equation}
    \dot{e}(r,t)=\mathcal{C}\dot{e}_\mathrm{S}\phi(r/r_\mathrm{S}),
\end{equation}
where $k$ and $\phi$ are functions of order unity and we suggest the constants $\mathcal{A}=0.5$ and $\mathcal{C}=-0.2$. The mass source term is included in the equation after the free expansion stage of CNM ($t\approx0.3\,\mathrm{kyr}$) and the energy source therm is added after $t_\mathrm{sf,CNM}\approx3\,\mathrm{kyr}$. The effects of different radial profiles (i.e., the shape of $k$ and $\phi$) remain to be investigated. Here we choose constant $k(r)=1$ and $\phi(r)=1$ motivated by \figu\ref{fig:evo_srcterms}. We implement this model and perform a 1D simulation using \athenapp{}.

Although we opt for a time-dependent solution for generality, we note that our simulation results are not very far from the self-similar evolution. As pointed out in \citet{White&Long1991ApJ...373..543W} (see Appendix \ref{app:snr1d} for further discussion of the details of the self-similar solution), when the SNR expansion is restricted to ST-like (i.e., $r_\mathrm{S}\propto t^{2/5}$), a similarity solution exists if $\dot{\rho}\propto t^{-1} k(r/r_\mathrm{S})$ and $\dot{e}\propto t^{-11/5}\phi(r/r_\mathrm{S})$. Or equivalently, we require the integrated properties $\dot{M}\propto \dot{M}_\mathrm{S}$ and $\dot{E}\propto \dot{E}_\mathrm{S}$. The SNR evolution shown in \figu\ref{fig:evo_fidu} indicates that, though there are significant differences with considerable energy loss before $t_\mathrm{sf,WNM}$, the expansion law is still ST-like. Our measurements of the source terms in \figu\ref{fig:evo_srcterms} hence suggest self-similar behaviors. We also note that the measured $\dot{\rho}$ here is significantly different from the work by \citet{Pittard2019MNRAS.488.3376P}, which assumed a spatially and temporally constant mass loading rate. Instead, our simulation results prefer the \citet{White&Long1991ApJ...373..543W} solution in the sense that the mass loading rate is time-dependent and scales as $t^{-1}$. 

In \figu\ref{fig:rad_theory}, we compare the simulation with the prediction of our time-dependent model at 10 kyr as well as other 1D models, the ST, WL \citep{White&Long1991ApJ...373..543W}, and P19 \citep{Pittard2019MNRAS.488.3376P} models. In the WL model, $\dot{\rho}=\mu_\mathrm{cl}/t_\mathrm{evap}(P/P_s)^{5/6}$ is assumed, where $\mu_\mathrm{cl}\equiv M_\mathrm{CNM}/M_\mathrm{WNM}$ (they denoted as $C$) and $t_\mathrm{evap}\equiv \tau t$. Given initial conditions, the parameter $\mu_\mathrm{cl}$ is already known ($\mu_\mathrm{cl}=10$ for our simulations). We use $\tau=5$ to better match the adiabatic results. In the P19 model, a spatially and temporally constant mass loading rate $\dot{\rho}=f_\mathrm{ML}\dot{\rho}_{\mathrm{S}, t=t_\mathrm{sf, WNM}}$ is assumed, where $f_\mathrm{ML}$ is a free parameter controlling the relative dominance of mass-loading. We show the P19 model with $f_\mathrm{ML}=10$ to match the adiabatic models. Note that neither 1D model considers any sink term for energy. 

Given that there is an additional mass loading term, all models predict higher interior density than the ST solution. However, in the models without energy loss term (WL and P19), all added mass also adds thermal energy and hence increases interior pressure significantly. In contrast, our model with energy loss term results in lower pressure than those models, while the temperature of all models is similar to each other. Qualitatively, the WL and P19 results are more consistent with the adiabatic simulations, overestimating pressure than our fiducial simulations with cooling. We also note that the size of 1D models is smaller than $r_S$ and similar to the radius of simulated SNRs at which the hot gas fraction drops to $f_V\sim0.5$. 

\section{Discussion and Summary}
\label{sec:summary}

In this work, we conduct high resolution (up to $1/64\,\mathrm{pc}$) hydrodynamical simulations of the evolution of supernova remnants in a cloudy medium realized by the saturation of thermal instability with an initial mean density of $10\mathrm{cm}^{-3}$. In our fiducial models, the background medium has the two-phase medium with the WNM at a density of $0.86\mathrm{cm}^{-3}$ and the CNM at a density of $84\mathrm{cm}^{-3}$ with a volume filling factor of $\sim 10\%$. We focus on the evolution of the hot, volume-filling component of SNRs during the nominal energy conserving stage (or Sedov-Taylor stage) of the WNM $t<t_{\rm sf,WNM}\sim 30\mathrm{kyr}$ during which the SNR builds up the hot gas mass and radial momentum, considering the effects of the shock-cloud interactions as impurities. 

Overall, the cloudy background affects the SNR evolution by loading more mass from clouds to the hot gas and sapping energy in the shock-cloud interfaces. This leads to the increase of the interior hot gas density and a more rapid reduction in energy. Thermal conduction enhances this effect but does not alter the global SNR evolution much in contrast to what is predicted in \citet{McKee&Ostriker1977ApJ...218..148M}. The SNR expansion slows down a bit, but still effectively follows the ST solution $r\propto t^{0.4}$. At $t_\mathrm{sf, WNM}$, the SNR lost significant energy through cooling in the shock-cloud interfaces, resulting in the absence of the PDS stage. The peak hot gas mass and radial momentum in a cloudy medium are $\sim 200 M_\odot$ and $\sim 10^5 M_\odot\mathrm{km/s}$, which are $\sim 50\%$ of the expected mass $M_{\rm sf}=410M_\odot$ and $\sim 80-90\%$ of the momentum $p_{\rm sf}=1.1\times10^5M_\odot\mathrm{km/s}$, respectively, at this time for a uniform medium with the same mean density~\citep{Kim2015ApJ...802...99K}.

The main conclusions are summarized below.

\begin{enumerate}
    \item The cloud ablation and evaporation do not alter the expansion law in the SNR evolution, especially the ST stage. However, the energetics changes significantly. Mass loading has been incorporated in previous effective models, \citep[e.g.,][]{White&Long1991ApJ...373..543W, Pittard2019MNRAS.488.3376P}. Here we find that the mass loading rate evolves with time with $\dot{\rho}\propto \dot{\rho}_\mathrm{S}\propto t^{-1}$. In addition, the energy sink is also important with a spatially flat cooling rate $\dot{e}\propto \dot{e}_\mathrm{S}\propto t^{-11/5}$. This motivates that a more realistic 1D model should include both mass loading and energy sink as source terms. As an illustration, we construct a minimal 1D model including both terms and show that it can qualitatively describe the evolution and internal structure of the SNR.
    \item Thermal conduction does not drastically change the overall expansion law and evolution of key global properties. However, observational properties of SNRs can be affected significantly by thermal conduction. The inclusion of thermal conduction smooths the internal structure of the hot gas, increases the density of hot gas in the central region increases significantly, and decreases the temperature. This leads to increased cooling rates, especially in regions surrounding the cold clouds and within the turbulent mixing zones. However, even though our treatment of thermal conduction is maximal as we neglect magnetic fields and anisotropic conduction, we note that the total cooling rate is still dominated by turbulent mixing. Additional conductive evaporation from cold clouds increases the contribution of the cold gas component to the hot gas phase. The modeling of X-ray from SNRs thus requires a careful treatment of thermal conduction.
    \item The results illustrate the complex internal structures of SNR, which is of particular interest due to its importance and uncertainty in constituting the third phase of ISM and determining the X-ray emission of the supernova remnants \citep{Zhang2019MNRAS.482.1602Z}. It may help to understand the mixed-morphology SNR \citep{Rho1994ApJ...430..757R}. In addition, \citet{Diesing2024arXiv240415396D} proposed an independent observational signal of shell formation arising from the interaction between nonthermal particles accelerated by the SNR forward shock (cosmic rays) and the dense shell formed during the radiative stage. The onset of the shell formation leads to much larger density enhancement compared to that from adiabatic shock, resulting in radio and $\gamma$-ray brightening by nearly two orders of magnitude. The complexity arising from the shock cloud interaction may or may not help such processes by producing ubiquitous density enhancements in the shock propagating into the CNM or ceasing the shock earlier by additional cooling at interfaces. Future direct MHD simulations with similar setups will shed light on this question.
\end{enumerate}

These findings improve our understanding of the diverse behaviors and characteristics exhibited by supernova remnants in realistic astrophysical environments, highlighting the importance of enhanced cooling in a cloudy medium.
A variety of physics is not considered including magnetic fields which can cause anisotropic conduction, nonequilibrium ionization, ambient cosmic rays, as well as multiple supernovae in which the explosion of the first supernova creates regions filled with hot gas leading to easier propagation of the subsequent explosion. The initial condition we adopt is not 100\% realistic and the SNR evolution can be sensitive to the distribution of clouds. In our simulations, the distribution of the cold clouds deviates from the picture of multiple discrete, isolated clouds scattered throughout the medium. For more realistic initial conditions, we can take galactic scale simulations like TIGRESS~\citep{Kim2017ApJ...846..133K} and resimulate SNR evolution within it, which we defer as future work.

\section*{Acknowledgments}
M.G. would like to thank Eliot Quataert, Eve Ostriker, Chris McKee, and Patrick Mullen for helpful discussions during the development of this work. The work of C.-G.K. was supported by NASA ATP grant No. 80NSSC22K0717. The authors are pleased to acknowledge that the work reported on in this paper was substantially performed using the Princeton Research Computing resources at Princeton University, which is consortium of groups led by the Princeton Institute for Computational Science and Engineering (PICSciE) and Office of Information Technology's Research Computing.

\vspace{5mm}

\software{\athenak{} \citep{Stone2024arXiv240916053S},}

\appendix

\begin{figure*}[htb]
    \centering
    \includegraphics[width=\linewidth]{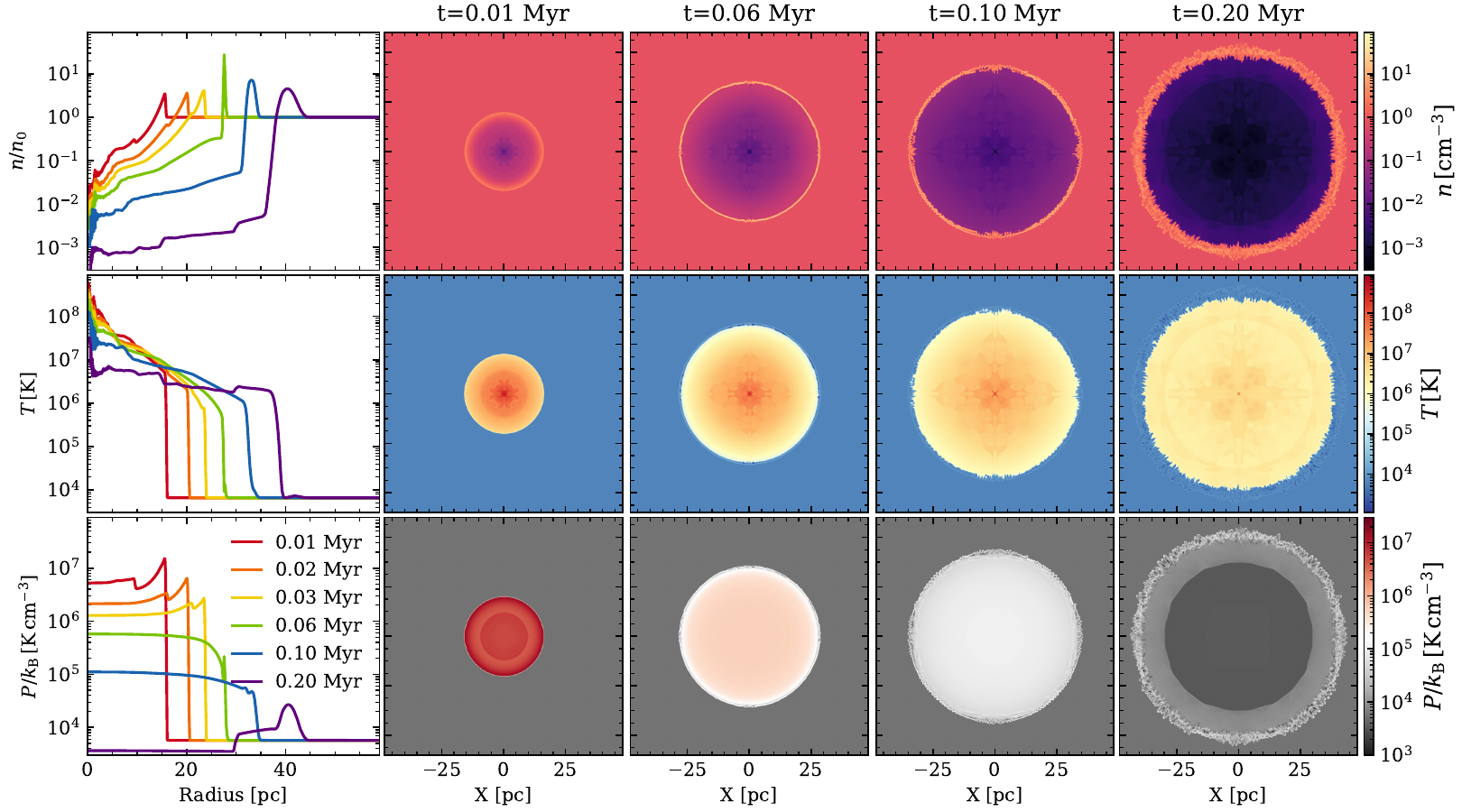}
    \caption{Radial profiles (left) and slice through the $z=0$ plane (right) of number density (top), temperature (middle), and pressure (bottom) of the SNR in a uniform background. The resolution is $\Delta x=1/8\,\mathrm{pc}$. There are only tiny asymmetries due to the Cartesian mesh.}
    \label{fig:snr_uni}
\end{figure*}

\section{Evolution of single supernova in uniform medium}\label{app:snr_uni}

As a reference, in \figu\ref{fig:snr_uni} we present the evolution of supernova remnant in a uniform background (model UN-1024 in \tab\ref{tab:models}) with the same properties as the WNM phase (density $n_0=0.86\,\mathrm{cm^{-3}}$ and temperature $T_0=6600\,\mathrm{K}$) in our fiducial run. Similar tests are performed in many works~\citep[e.g.,][]{Kim2015ApJ...802...99K, Walch2015MNRAS.451.2757W}. The simulation adopts a box of $1024^3$ cells with a resolution $\Delta x=1/8\,\mathrm{pc}$. A dense ($n/n_0\gtrsim 30$) cold thin shell forms when cooling becomes dominant ($t\gtrsim 0.03\,\mathrm{Myr}$) with the thin-shell instability arising. Cartesian mesh introduces small asymmetries in the development of the thin shell instability when the shock propagates along the grid direction at $\sim 0.1\,\mathrm{Myr}$ but the effect is relatively weak. 

\section{Evolution of supernova remnant with mass loading and cooling}\label{app:snr1d}
A general 1D-form of the evolution of supernova remnants is 
\begin{align}
    \frac{\partial \rho}{\partial t} + \frac{1}{r^2}\frac{\partial}{\partial r}(r^2 v\rho)&=\dot{\rho},\\
    \frac{\partial \rho v}{\partial t} + \frac{1}{r^2}\frac{\partial}{\partial r}(r^2\rho v^2) + \frac{\partial P}{\partial r}&=\dot{\varrho},\\
    \frac{\partial \mathcal{E}}{\partial t} + \frac{1}{r^2}\frac{\partial}{\partial r}\left[r^2\rho v\left(\frac{1}{2}v^2+\frac{\gamma}{\gamma-1}\frac{P}{\rho}\right)\right]&=\dot{e},
\end{align}
where $\rho$, $v$, $P$, and $\mathcal{E}=e+\rho v^2/2=P/(\gamma-1)+\rho v^2/2$ are the density, velocity, pressure, and total energy density. The additional terms, $\dot{\rho}$, $\dot{\varrho}$, and $\dot{e}$, are the source or sink terms for mass, momentum, and energy.
We can rewrite them using the form of the velocity and entropy equations:
\begin{align}
    \frac{\partial \rho}{\partial t} + \frac{1}{r^2}\frac{\partial}{\partial r}(r^2 v\rho)&=\dot{\rho},\\
    \frac{\partial v}{\partial t} + v\frac{\partial v}{\partial r} + \frac{1}{\rho} \frac{\partial P}{\partial r}&=\frac{\dot{\varrho}-\dot{\rho}v}{\rho},\\
    \frac{P}{\gamma-1}\frac{1}{S}\frac{d S}{d t}&=\dot{e}-\dot{\varrho}v+\dot{\rho}\left(\frac{1}{2}v^2-\frac{\gamma}{\gamma-1}\frac{P}{\rho}\right),
\end{align}
where $S\equiv P/\rho^\gamma$. 

To find a self-similarity solution, we make the following substitutions:
\begin{equation}
\begin{aligned}
    P=P_sf(x),
    \rho=\rho_sg(x),
    v=v_sh(x),\\
    \dot{\rho}=\dot{\rho}_sk(x),
    \dot{\varrho}=\dot{\varrho}_s\phi(x),
    \dot{e}=\dot{e}_s\psi(x),
\end{aligned}
\end{equation}
where $x\equiv r/r_s$ and $\rho_s\equiv\rho_0(\gamma+1)/(\gamma-1)$ with $f(1)=g(1)=k(1)=\phi(1)=\psi(1)=1$ and $h(1)=2/(\gamma+1)$. As for the usual ST solution, the dependence of $P_s$, $v_s$, and $\rho_s$ on $r_s$ is determined by the requirement that the $r_s$ dependence drop out of the resulting ordinary differential equations:
\begin{align}
    P_s&=\frac{KE}{2\pi r_s^3},\\
    v_s&=\sqrt{\frac{(\gamma+1)KE}{4\pi\rho_0r_s^3}},\\
    r_s&=\left[\frac{25(\gamma+1)KE}{16\pi\rho_0}\right]^{1/5}t^{2/5},
\end{align}
where $E$ is the energy of the SN explosion and
\begin{align}
    K&=(\gamma-1)\bigg/\int_0^1x^2dx\left[2f+\frac{1}{2}(\gamma+1)gh^2\right],
\end{align}
is a dimensionless constant that determines the ratio of thermal to kinetic energy in the SNR. The scaling of all these variables is exactly the same as for the standard ST solution without evaporating clouds, but the value of the constant $K$ and the energy $E$ differs when evaporation and additional cooling are included.

Then we aim to solve the functions $f(x)$, $g(x)$, and $h(x)$ given functions $k(x)$, $\phi(x)$, and $\psi(x)$. The resulting ordinary differential equations for $f$, $g$, and $h$ are
\begin{align}
    (h-x)g'+gh'=-\frac{2gh}{x}+Ak,&\\
    \frac{2(\gamma-1)}{(\gamma+1)^2}f'+g(h-x)h'=\frac{3}{2}gh+B\phi-Akh,&\\
    \begin{aligned}
        g(h-x)f'-\gamma f(h-x)g'=
        3fg+&\\ 
        \frac{(\gamma+1)^2}{2}g\left(C\psi-B\phi h\right)
        +&\\
        Ak\left[\frac{(\gamma+1)^2}{4}gh^2-\gamma f\right]&,
    \end{aligned}&
\end{align}
where
\begin{equation}
\begin{aligned}
    A\equiv\dot{\rho}_sr_s/v_s\rho_s,
    B\equiv\dot{\varrho}_sr_s/v_s^2\rho_s,
    C\equiv\dot{e}_sr_s/v_s^3\rho_s.
\end{aligned}    
\end{equation}
For a similarity solution to exist, $A$, $B$, and $C$ must be constant. This implies that
\begin{equation}
\begin{aligned}
    \dot{\rho}_s\propto v_s/r_s \propto t^{-1},\\
    \dot{\varrho}_s\propto v_s^2/r_s \propto t^{-8/5},\\
    \dot{e}_s\propto v_s^3/r_s \propto t^{-11/5},
\end{aligned}    
\end{equation}
or equivalently for integrated variables,
\begin{equation}
\begin{aligned}
    \dot{M} \propto \dot{\rho}_s r_s^3 \propto v_s r_s^2\propto t^{1/5},\\ 
    \dot{p} \propto \dot{\varrho}_s r_s^3 \propto v_s^2r_s^2 \propto t^{-2/5}, \\ 
    \dot{E} \propto \dot{e}_s r_s^3 \propto v_s^3r_s^2\propto t^{-1}.
\end{aligned}    
\end{equation}
Thus if a self-similar solution exists, mass loading, momentum loading, and energy loading will follow a certain scaling of time, i.e., $\dot{\rho}_s\propto t^{-1}, \dot{\varrho}_s \propto t^{-8/5}, \dot{e}_s \propto t^{-11/5}$, and correspondingly, $\dot{M}\propto t^{1/5}$, $\dot{p}\propto t^{-2/5}$, and $\dot{E}\propto t^{-1}$.

\bibliography{main}{}

\begin{thebibliography}{}
\expandafter\ifx\csname natexlab\endcsname\relax\def\natexlab#1{#1}\fi
\providecommand{\url}[1]{\href{#1}{#1}}
\providecommand{\dodoi}[1]{doi:~\href{http://doi.org/#1}{\nolinkurl{#1}}}
\providecommand{\doeprint}[1]{\href{http://ascl.net/#1}{\nolinkurl{http://ascl.net/#1}}}
\providecommand{\doarXiv}[1]{\href{https://arxiv.org/abs/#1}{\nolinkurl{https://arxiv.org/abs/#1}}}

\bibitem[{{Abruzzo} {et~al.}(2022){Abruzzo}, {Bryan}, \&
  {Fielding}}]{Abruzzo2022ApJ...925..199A}
{Abruzzo}, M.~W., {Bryan}, G.~L., \& {Fielding}, D.~B. 2022, \apj, 925, 199,
  \dodoi{10.3847/1538-4357/ac3c48}

\bibitem[{{Armillotta} {et~al.}(2016){Armillotta}, {Fraternali}, \&
  {Marinacci}}]{Armillotta2016MNRAS.462.4157A}
{Armillotta}, L., {Fraternali}, F., \& {Marinacci}, F. 2016, \mnras, 462, 4157,
  \dodoi{10.1093/mnras/stw1930}

\bibitem[{{Banda-Barrag{\'a}n} {et~al.}(2020){Banda-Barrag{\'a}n},
  {Br{\"u}ggen}, {Federrath}, {Wagner}, {Scannapieco}, \&
  {Cottle}}]{2020MNRAS.499.2173B}
{Banda-Barrag{\'a}n}, W.~E., {Br{\"u}ggen}, M., {Federrath}, C., {et~al.} 2020,
  \mnras, 499, 2173, \dodoi{10.1093/mnras/staa2904}

\bibitem[{{Bedogni} \& {Woodward}(1990)}]{Bedogni1990A&A...231..481B}
{Bedogni}, R., \& {Woodward}, P.~R. 1990, \aap, 231, 481

\bibitem[{{Cioffi} {et~al.}(1988){Cioffi}, {McKee}, \&
  {Bertschinger}}]{Cioffi1988ApJ...334..252C}
{Cioffi}, D.~F., {McKee}, C.~F., \& {Bertschinger}, E. 1988, \apj, 334, 252,
  \dodoi{10.1086/166834}

\bibitem[{{Cowie} \& {McKee}(1977)}]{Cowie&McKee1977ApJ...211..135C}
{Cowie}, L.~L., \& {McKee}, C.~F. 1977, \apj, 211, 135, \dodoi{10.1086/154911}

\bibitem[{{Cowie} {et~al.}(1981){Cowie}, {McKee}, \&
  {Ostriker}}]{Cowie1981ApJ...247..908C}
{Cowie}, L.~L., {McKee}, C.~F., \& {Ostriker}, J.~P. 1981, \apj, 247, 908,
  \dodoi{10.1086/159100}

\bibitem[{{Cox}(1972)}]{Cox1972ApJ...178..159C}
{Cox}, D.~P. 1972, \apj, 178, 159, \dodoi{10.1086/151775}

\bibitem[{{Cox}(2005)}]{Cox2005ARA&A..43..337C}
---. 2005, \araa, 43, 337, \dodoi{10.1146/annurev.astro.43.072103.150615}

\bibitem[{{Cox} \& {Smith}(1974)}]{Cox&Smith1974ApJ...189L.105C}
{Cox}, D.~P., \& {Smith}, B.~W. 1974, \apjl, 189, L105, \dodoi{10.1086/181476}

\bibitem[{{Diesing} {et~al.}(2024){Diesing}, {Guo}, {Kim}, {Stone}, \&
  {Caprioli}}]{Diesing2024arXiv240415396D}
{Diesing}, R., {Guo}, M., {Kim}, C.-G., {Stone}, J., \& {Caprioli}, D. 2024,
  arXiv e-prints, arXiv:2404.15396, \dodoi{10.48550/arXiv.2404.15396}

\bibitem[{{Draine}(2011)}]{Draine2011piim.book.....D}
{Draine}, B.~T. 2011, {Physics of the Interstellar and Intergalactic Medium}

\bibitem[{{Dyson} {et~al.}(2002){Dyson}, {Arthur}, \&
  {Hartquist}}]{Dyson2002A&A...390.1063D}
{Dyson}, J.~E., {Arthur}, S.~J., \& {Hartquist}, T.~W. 2002, \aap, 390, 1063,
  \dodoi{10.1051/0004-6361:20020731}

\bibitem[{{El-Badry} {et~al.}(2019){El-Badry}, {Ostriker}, {Kim}, {Quataert},
  \& {Weisz}}]{El-Badry2019MNRAS.490.1961E}
{El-Badry}, K., {Ostriker}, E.~C., {Kim}, C.-G., {Quataert}, E., \& {Weisz},
  D.~R. 2019, \mnras, 490, 1961, \dodoi{10.1093/mnras/stz2773}

\bibitem[{{Field}(1965)}]{Field1965ApJ...142..531F}
{Field}, G.~B. 1965, \apj, 142, 531, \dodoi{10.1086/148317}

\bibitem[{{Field} {et~al.}(1969){Field}, {Goldsmith}, \&
  {Habing}}]{Field1969ApJ...155L.149F}
{Field}, G.~B., {Goldsmith}, D.~W., \& {Habing}, H.~J. 1969, \apjl, 155, L149,
  \dodoi{10.1086/180324}

\bibitem[{{Fielding} \& {Bryan}(2022)}]{Fielding2022ApJ...924...82F}
{Fielding}, D.~B., \& {Bryan}, G.~L. 2022, \apj, 924, 82,
  \dodoi{10.3847/1538-4357/ac2f41}

\bibitem[{{Fielding} {et~al.}(2020){Fielding}, {Ostriker}, {Bryan}, \&
  {Jermyn}}]{Fielding2020ApJ...894L..24F}
{Fielding}, D.~B., {Ostriker}, E.~C., {Bryan}, G.~L., \& {Jermyn}, A.~S. 2020,
  \apjl, 894, L24, \dodoi{10.3847/2041-8213/ab8d2c}

\bibitem[{{Gazol} \& {Kim}(2013)}]{Gazol2013ApJ...765...49G}
{Gazol}, A., \& {Kim}, J. 2013, \apj, 765, 49,
  \dodoi{10.1088/0004-637X/765/1/49}

\bibitem[{{Goldsmith} \& {Pittard}(2017)}]{Goldsmith2017MNRAS.470.2427G}
{Goldsmith}, K.~J.~A., \& {Pittard}, J.~M. 2017, \mnras, 470, 2427,
  \dodoi{10.1093/mnras/stx1431}

\bibitem[{{Gregori} {et~al.}(1999){Gregori}, {Miniati}, {Ryu}, \&
  {Jones}}]{Gregori1999ApJ...527L.113G}
{Gregori}, G., {Miniati}, F., {Ryu}, D., \& {Jones}, T.~W. 1999, \apjl, 527,
  L113, \dodoi{10.1086/312402}

\bibitem[{{Gronke} \& {Oh}(2018)}]{Gronke2018MNRAS.480L.111G}
{Gronke}, M., \& {Oh}, S.~P. 2018, \mnras, 480, L111,
  \dodoi{10.1093/mnrasl/sly131}

\bibitem[{{Gronke} {et~al.}(2022){Gronke}, {Oh}, {Ji}, \&
  {Norman}}]{Gronke2022MNRAS.511..859G}
{Gronke}, M., {Oh}, S.~P., {Ji}, S., \& {Norman}, C. 2022, \mnras, 511, 859,
  \dodoi{10.1093/mnras/stab3351}

\bibitem[{{Ho} {et~al.}(2024){Ho}, {Yuen}, \&
  {Lazarian}}]{Ho2024arXiv240714199H}
{Ho}, K.~W., {Yuen}, K.~H., \& {Lazarian}, A. 2024, arXiv e-prints,
  arXiv:2407.14199, \dodoi{10.48550/arXiv.2407.14199}

\bibitem[{{Iffrig} \& {Hennebelle}(2015)}]{Iffrig2015A&A...576A..95I}
{Iffrig}, O., \& {Hennebelle}, P. 2015, \aap, 576, A95,
  \dodoi{10.1051/0004-6361/201424556}

\bibitem[{{Kim} {et~al.}(2023){Kim}, {Kim}, {Gong}, \&
  {Ostriker}}]{Kim2023ApJ...946....3K}
{Kim}, C.-G., {Kim}, J.-G., {Gong}, M., \& {Ostriker}, E.~C. 2023, \apj, 946,
  3, \dodoi{10.3847/1538-4357/acbd3a}

\bibitem[{{Kim} \& {Ostriker}(2015)}]{Kim2015ApJ...802...99K}
{Kim}, C.-G., \& {Ostriker}, E.~C. 2015, \apj, 802, 99,
  \dodoi{10.1088/0004-637X/802/2/99}

\bibitem[{{Kim} \& {Ostriker}(2017)}]{Kim2017ApJ...846..133K}
---. 2017, \apj, 846, 133, \dodoi{10.3847/1538-4357/aa8599}

\bibitem[{{Kim} {et~al.}(2017){Kim}, {Ostriker}, \&
  {Raileanu}}]{Kim2017ApJ...834...25K}
{Kim}, C.-G., {Ostriker}, E.~C., \& {Raileanu}, R. 2017, \apj, 834, 25,
  \dodoi{10.3847/1538-4357/834/1/25}

\bibitem[{{Kim} {et~al.}(2020){Kim}, {Ostriker}, {Somerville}, {Bryan},
  {Fielding}, {Forbes}, {Hayward}, {Hernquist}, \&
  {Pandya}}]{Kim2020ApJ...900...61K}
{Kim}, C.-G., {Ostriker}, E.~C., {Somerville}, R.~S., {et~al.} 2020, \apj, 900,
  61, \dodoi{10.3847/1538-4357/aba962}

\bibitem[{{Klein} {et~al.}(1994){Klein}, {McKee}, \&
  {Colella}}]{Klein1994ApJ...420..213K}
{Klein}, R.~I., {McKee}, C.~F., \& {Colella}, P. 1994, \apj, 420, 213,
  \dodoi{10.1086/173554}

\bibitem[{{Koyama} \& {Inutsuka}(2002)}]{Koyama2002ApJ...564L..97K}
{Koyama}, H., \& {Inutsuka}, S.-i. 2002, \apjl, 564, L97,
  \dodoi{10.1086/338978}

\bibitem[{{Lancaster} {et~al.}(2024){Lancaster}, {Ostriker}, {Kim}, {Kim}, \&
  {Bryan}}]{Lancaster2024ApJ...970...18L}
{Lancaster}, L., {Ostriker}, E.~C., {Kim}, C.-G., {Kim}, J.-G., \& {Bryan},
  G.~L. 2024, \apj, 970, 18, \dodoi{10.3847/1538-4357/ad47f6}

\bibitem[{{Lancaster} {et~al.}(2021{\natexlab{a}}){Lancaster}, {Ostriker},
  {Kim}, \& {Kim}}]{Lancaster2021ApJ...914...89L}
{Lancaster}, L., {Ostriker}, E.~C., {Kim}, J.-G., \& {Kim}, C.-G.
  2021{\natexlab{a}}, \apj, 914, 89, \dodoi{10.3847/1538-4357/abf8ab}

\bibitem[{{Lancaster} {et~al.}(2021{\natexlab{b}}){Lancaster}, {Ostriker},
  {Kim}, \& {Kim}}]{Lancaster2021ApJ...914...90L}
---. 2021{\natexlab{b}}, \apj, 914, 90, \dodoi{10.3847/1538-4357/abf8ac}

\bibitem[{{Lemaster} \& {Stone}(2009)}]{Lemaster&Stone2009ApJ...691.1092L}
{Lemaster}, M.~N., \& {Stone}, J.~M. 2009, \apj, 691, 1092,
  \dodoi{10.1088/0004-637X/691/2/1092}

\bibitem[{{Li} {et~al.}(2015){Li}, {Ostriker}, {Cen}, {Bryan}, \&
  {Naab}}]{Li2015ApJ...814....4L}
{Li}, M., {Ostriker}, J.~P., {Cen}, R., {Bryan}, G.~L., \& {Naab}, T. 2015,
  \apj, 814, 4, \dodoi{10.1088/0004-637X/814/1/4}

\bibitem[{{Li} {et~al.}(2020){Li}, {Hopkins}, {Squire}, \&
  {Hummels}}]{Li2020MNRAS.492.1841L}
{Li}, Z., {Hopkins}, P.~F., {Squire}, J., \& {Hummels}, C. 2020, \mnras, 492,
  1841, \dodoi{10.1093/mnras/stz3567}

\bibitem[{{Mac Low} {et~al.}(1994){Mac Low}, {McKee}, {Klein}, {Stone}, \&
  {Norman}}]{MacLow1994ApJ...433..757M}
{Mac Low}, M.-M., {McKee}, C.~F., {Klein}, R.~I., {Stone}, J.~M., \& {Norman},
  M.~L. 1994, \apj, 433, 757, \dodoi{10.1086/174685}

\bibitem[{{Martizzi} {et~al.}(2015){Martizzi}, {Faucher-Gigu{\`e}re}, \&
  {Quataert}}]{Martizzi2015MNRAS.450..504M}
{Martizzi}, D., {Faucher-Gigu{\`e}re}, C.-A., \& {Quataert}, E. 2015, \mnras,
  450, 504, \dodoi{10.1093/mnras/stv562}

\bibitem[{{McKee} \& {Ostriker}(1977)}]{McKee&Ostriker1977ApJ...218..148M}
{McKee}, C.~F., \& {Ostriker}, J.~P. 1977, \apj, 218, 148,
  \dodoi{10.1086/155667}

\bibitem[{{Orlando} {et~al.}(2008){Orlando}, {Bocchino}, {Reale}, {Peres}, \&
  {Pagano}}]{Orlando2008ApJ...678..274O}
{Orlando}, S., {Bocchino}, F., {Reale}, F., {Peres}, G., \& {Pagano}, P. 2008,
  \apj, 678, 274, \dodoi{10.1086/529420}

\bibitem[{{Orlando} {et~al.}(2005){Orlando}, {Peres}, {Reale}, {Bocchino},
  {Rosner}, {Plewa}, \& {Siegel}}]{Orlando2005A&A...444..505O}
{Orlando}, S., {Peres}, G., {Reale}, F., {et~al.} 2005, \aap, 444, 505,
  \dodoi{10.1051/0004-6361:20052896}

\bibitem[{{Ostriker} {et~al.}(2001){Ostriker}, {Stone}, \&
  {Gammie}}]{Ostriker2001ApJ...546..980O}
{Ostriker}, E.~C., {Stone}, J.~M., \& {Gammie}, C.~F. 2001, \apj, 546, 980,
  \dodoi{10.1086/318290}

\bibitem[{{Ostriker} \& {McKee}(1988)}]{Ostriker&McKee1988RvMP...60....1O}
{Ostriker}, J.~P., \& {McKee}, C.~F. 1988, Reviews of Modern Physics, 60, 1,
  \dodoi{10.1103/RevModPhys.60.1}

\bibitem[{{Parker}(1953)}]{Parker1953ApJ...117..431P}
{Parker}, E.~N. 1953, \apj, 117, 431, \dodoi{10.1086/145707}

\bibitem[{{Pittard}(2019)}]{Pittard2019MNRAS.488.3376P}
{Pittard}, J.~M. 2019, \mnras, 488, 3376, \dodoi{10.1093/mnras/stz1885}

\bibitem[{{Pittard} {et~al.}(2003){Pittard}, {Arthur}, {Dyson}, {Falle},
  {Hartquist}, {Knight}, \& {Pexton}}]{Pittard2003A&A...401.1027P}
{Pittard}, J.~M., {Arthur}, S.~J., {Dyson}, J.~E., {et~al.} 2003, \aap, 401,
  1027, \dodoi{10.1051/0004-6361:20030157}

\bibitem[{{Pittard} \& {Parkin}(2016)}]{Pittard&Parkin2016MNRAS.457.4470P}
{Pittard}, J.~M., \& {Parkin}, E.~R. 2016, \mnras, 457, 4470,
  \dodoi{10.1093/mnras/stw025}

\bibitem[{{Rathjen} {et~al.}(2021){Rathjen}, {Naab}, {Girichidis}, {Walch},
  {W{\"u}nsch}, {Dinnbier}, {Seifried}, {Klessen}, \&
  {Glover}}]{Rathjen2021MNRAS.504.1039R}
{Rathjen}, T.-E., {Naab}, T., {Girichidis}, P., {et~al.} 2021, \mnras, 504,
  1039, \dodoi{10.1093/mnras/stab900}

\bibitem[{{Rho} {et~al.}(1994){Rho}, {Petre}, {Schlegel}, \&
  {Hester}}]{Rho1994ApJ...430..757R}
{Rho}, J., {Petre}, R., {Schlegel}, E.~M., \& {Hester}, J.~J. 1994, \apj, 430,
  757, \dodoi{10.1086/174446}

\bibitem[{{Schneider} \& {Robertson}(2018)}]{Schneider2018ApJ...860..135S}
{Schneider}, E.~E., \& {Robertson}, B.~E. 2018, \apj, 860, 135,
  \dodoi{10.3847/1538-4357/aac329}

\bibitem[{{Schure} {et~al.}(2009){Schure}, {Kosenko}, {Kaastra}, {Keppens}, \&
  {Vink}}]{Schure2009A&A...508..751S}
{Schure}, K.~M., {Kosenko}, D., {Kaastra}, J.~S., {Keppens}, R., \& {Vink}, J.
  2009, \aap, 508, 751, \dodoi{10.1051/0004-6361/200912495}

\bibitem[{{Sedov}(1959)}]{Sedov1959sdmm.book.....S}
{Sedov}, L.~I. 1959, {Similarity and Dimensional Methods in Mechanics}

\bibitem[{{Shin} {et~al.}(2008){Shin}, {Stone}, \&
  {Snyder}}]{Shin2008ApJ...680..336S}
{Shin}, M.-S., {Stone}, J.~M., \& {Snyder}, G.~F. 2008, \apj, 680, 336,
  \dodoi{10.1086/587775}

\bibitem[{{Slavin} {et~al.}(2017){Slavin}, {Smith}, {Foster}, {Winter},
  {Raymond}, {Slane}, \& {Yamaguchi}}]{Slavin2017ApJ...846...77S}
{Slavin}, J.~D., {Smith}, R.~K., {Foster}, A., {et~al.} 2017, \apj, 846, 77,
  \dodoi{10.3847/1538-4357/aa8552}

\bibitem[{{Sparre} {et~al.}(2020){Sparre}, {Pfrommer}, \&
  {Ehlert}}]{Sparre2020MNRAS.499.4261S}
{Sparre}, M., {Pfrommer}, C., \& {Ehlert}, K. 2020, \mnras, 499, 4261,
  \dodoi{10.1093/mnras/staa3177}

\bibitem[{{Spitzer}(1962)}]{Spitzer1962pfig.book.....S}
{Spitzer}, L. 1962, {Physics of Fully Ionized Gases}

\bibitem[{{Stone} {et~al.}(2020){Stone}, {Tomida}, {White}, \&
  {Felker}}]{Stone2020ApJS..249....4S}
{Stone}, J.~M., {Tomida}, K., {White}, C.~J., \& {Felker}, K.~G. 2020, \apjs,
  249, 4, \dodoi{10.3847/1538-4365/ab929b}

\bibitem[{{Stone} {et~al.}(2024){Stone}, {Mullen}, {Fielding}, {Grete}, {Guo},
  {Kempski}, {Most}, {White}, \& {Wong}}]{Stone2024arXiv240916053S}
{Stone}, J.~M., {Mullen}, P.~D., {Fielding}, D., {et~al.} 2024, arXiv e-prints,
  arXiv:2409.16053, \dodoi{10.48550/arXiv.2409.16053}

\bibitem[{{Tan} \& {Fielding}(2024)}]{Tan2024MNRAS.527.9683T}
{Tan}, B., \& {Fielding}, D.~B. 2024, \mnras, 527, 9683,
  \dodoi{10.1093/mnras/stad3793}

\bibitem[{{Tan} {et~al.}(2021){Tan}, {Oh}, \&
  {Gronke}}]{Tan2021MNRAS.502.3179T}
{Tan}, B., {Oh}, S.~P., \& {Gronke}, M. 2021, \mnras, 502, 3179,
  \dodoi{10.1093/mnras/stab053}

\bibitem[{{Taylor}(1950)}]{Taylor1950RSPSA.201..159T}
{Taylor}, G. 1950, Proceedings of the Royal Society of London Series A, 201,
  159, \dodoi{10.1098/rspa.1950.0049}

\bibitem[{{Vink}(2012)}]{Vink2012A&ARv..20...49V}
{Vink}, J. 2012, \aapr, 20, 49, \dodoi{10.1007/s00159-011-0049-1}

\bibitem[{{Walch} \& {Naab}(2015)}]{Walch2015MNRAS.451.2757W}
{Walch}, S., \& {Naab}, T. 2015, \mnras, 451, 2757,
  \dodoi{10.1093/mnras/stv1155}

\bibitem[{{White} \& {Long}(1991)}]{White&Long1991ApJ...373..543W}
{White}, R.~L., \& {Long}, K.~S. 1991, \apj, 373, 543, \dodoi{10.1086/170073}

\bibitem[{{Wolfire} {et~al.}(1995){Wolfire}, {Hollenbach}, {McKee}, {Tielens},
  \& {Bakes}}]{Wolfire1995ApJ...443..152W}
{Wolfire}, M.~G., {Hollenbach}, D., {McKee}, C.~F., {Tielens}, A.~G.~G.~M., \&
  {Bakes}, E.~L.~O. 1995, \apj, 443, 152, \dodoi{10.1086/175510}

\bibitem[{{Woltjer}(1972)}]{Woltjer1972ARA&A..10..129W}
{Woltjer}, L. 1972, \araa, 10, 129, \dodoi{10.1146/annurev.aa.10.090172.001021}

\bibitem[{{Zhang} \& {Chevalier}(2019)}]{Zhang2019MNRAS.482.1602Z}
{Zhang}, D., \& {Chevalier}, R.~A. 2019, \mnras, 482, 1602,
  \dodoi{10.1093/mnras/sty2769}

\end{thebibliography}
\bibliographystyle{aasjournal}

\end{CJK*}
\end{document}